\documentclass[acmtog]{acmart} 
\acmSubmissionID{329}

\usepackage{booktabs} 

\usepackage{multirow}
\usepackage{dsfont}
\usepackage{gensymb}

\usepackage{enumitem}

\usepackage[ruled]{algorithm2e} 

\SetAlFnt{\small}
\SetAlCapFnt{\small}
\SetAlCapNameFnt{\small}
\SetAlCapHSkip{0pt}

\acmJournal{TOG}
\acmYear{2025} 
\acmVolume{44} 
\acmNumber{4} 
\acmArticle{} 
\acmMonth{8} 
\acmPrice{}
\acmDOI{10.1145/3730836}

\setcopyright{rightsretained}

\begin{document}
\begin{teaserfigure}
    \includegraphics[width=\linewidth]{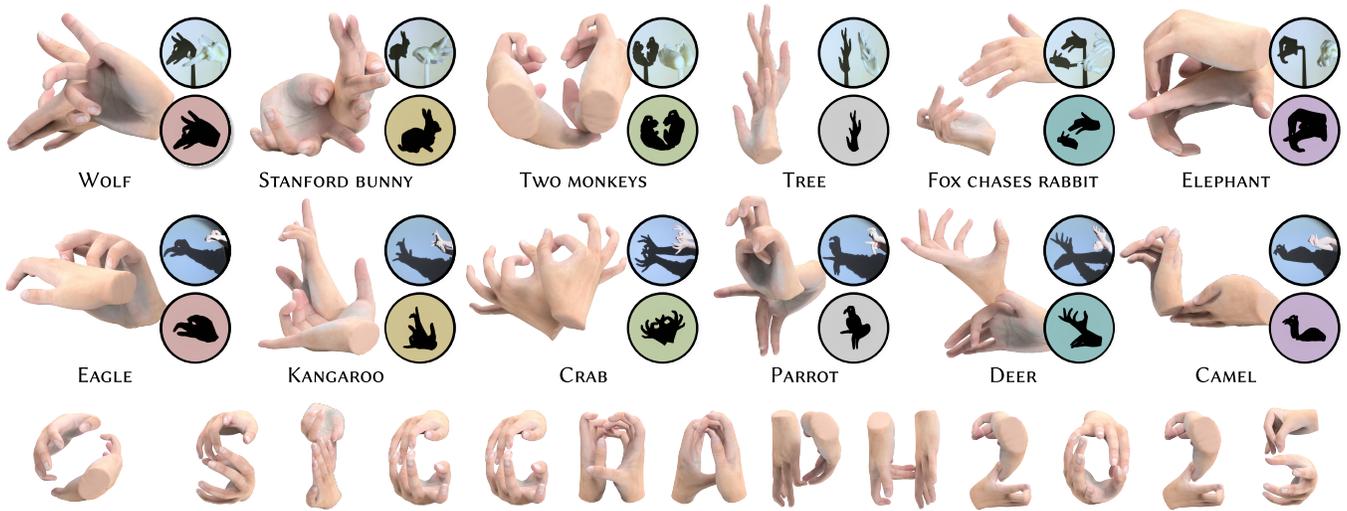}
    \vspace*{-5mm}
    \caption{
    3D poses of left and right hands reconstructed by our method for producing shadows of different target objects.
    Lower insets show renderings of each 3D hand-pose result with the viewpoint set at the light source location, thus essentially revealing the final shadows produced by the respective hand poses. 
    The upper insets in the first row show their 3D prints, whereas those in the second row show real shadows produced by human hands.
    }
    \Description[]{}
    \label{fig:teaser}
\end{teaserfigure}

\title{Hand-Shadow Poser}


\author[*]{Hao Xu}
\orcid{0000-0003-3676-5737}
\authornote{indicates joint first authors.}

\author{Yinqiao Wang}
\orcid{0000-0002-6099-206X}
\affiliation{%
    \institution{The Chinese University of Hong Kong}
    \country{Hong Kong, China}
}
\email{{xuhao, yqwang}@cse.cuhk.edu.hk}
\authornotemark[1]

\author{Niloy J. Mitra}
\orcid{0000-0002-2597-0914}
\affiliation{%
    \institution{University College London, Adobe Research}
    \city{London}
    \country{United Kingdom}
}
\email{n.mitra@cs.ucl.ac.uk}

\author{Shuaicheng Liu}
\orcid{0000-0002-8815-5335}
\affiliation{%
    \institution{University of Electronic Science and Technology of China}
    \city{Chengdu}
    \country{China}
}
\email{liushuaicheng@uestc.edu.cn}

\author[*]{Pheng-Ann Heng}
\orcid{0000-0003-3055-5034}

\author{Chi-Wing Fu}
\orcid{0000-0002-5238-593X}
\affiliation{%
    \institution{The Chinese University of Hong Kong}
    \country{Hong Kong, China}
}
\email{{pheng,cwfu}@cse.cuhk.edu.hk}

\renewcommand\shortauthors{Xu, H. et al.}

\begin{abstract}
Hand shadow art is a captivating art form, creatively using hand shadows to reproduce expressive shapes on the wall.
%
In this work, we study an inverse problem: given a target shape, find the poses of left and right hands that together best produce a shadow resembling the input.
%
This problem is nontrivial, since the design space of 3D hand poses is huge while being restrictive due to anatomical constraints.
Also, we need to attend to the input's shape and crucial features, though the input is colorless and textureless. 
%
To meet these challenges, we design Hand-Shadow Poser, a three-stage pipeline, to decouple the anatomical constraints (by hand) and semantic constraints (by shadow shape):
(i)~a generative hand assignment module to explore diverse but reasonable left/right-hand shape hypotheses;
(ii)~a generalized hand-shadow alignment module to infer coarse hand poses with a similarity-driven strategy for selecting hypotheses;
and (iii)~a shadow-feature-aware refinement module to optimize the hand poses for physical plausibility and shadow feature preservation.
%
Further, we design our pipeline to be trainable on generic public hand data, thus avoiding the need for any specialized training dataset. 
%
For method validation, we build a benchmark of 210 diverse shadow shapes of varying complexity and a comprehensive set of metrics, including a novel DINOv2-based evaluation metric.
%
Through extensive comparisons with multiple baselines and user studies, our approach is demonstrated to effectively generate bimanual hand poses for a large variety of hand shapes for over 85\% of the benchmark cases.
\end{abstract}

%
%


%
%

\makeatletter
\DeclareRobustCommand\onedot{\futurelet\@let@token\@onedot}
\def\@onedot{\ifx\@let@token.\else.\null\fi\xspace}

\def\eg{\emph{e.g}\onedot} \def\Eg{\emph{E.g}\onedot}
\def\ie{\emph{i.e}\onedot} \def\Ie{\emph{I.e}\onedot}
\def\cf{\emph{cf}\onedot} \def\Cf{\emph{Cf}\onedot}
\def\etc{\emph{etc}\onedot} \def\vs{\emph{vs}\onedot}
\def\wrt{w.r.t\onedot} \def\dof{d.o.f\onedot}
\def\iid{i.i.d\onedot} \def\wolog{w.l.o.g\onedot}
\def\etal{et al\onedot}
\makeatother


\maketitle

\section{Introduction}
\begin{figure}[t]
    \includegraphics[width=\linewidth]
    {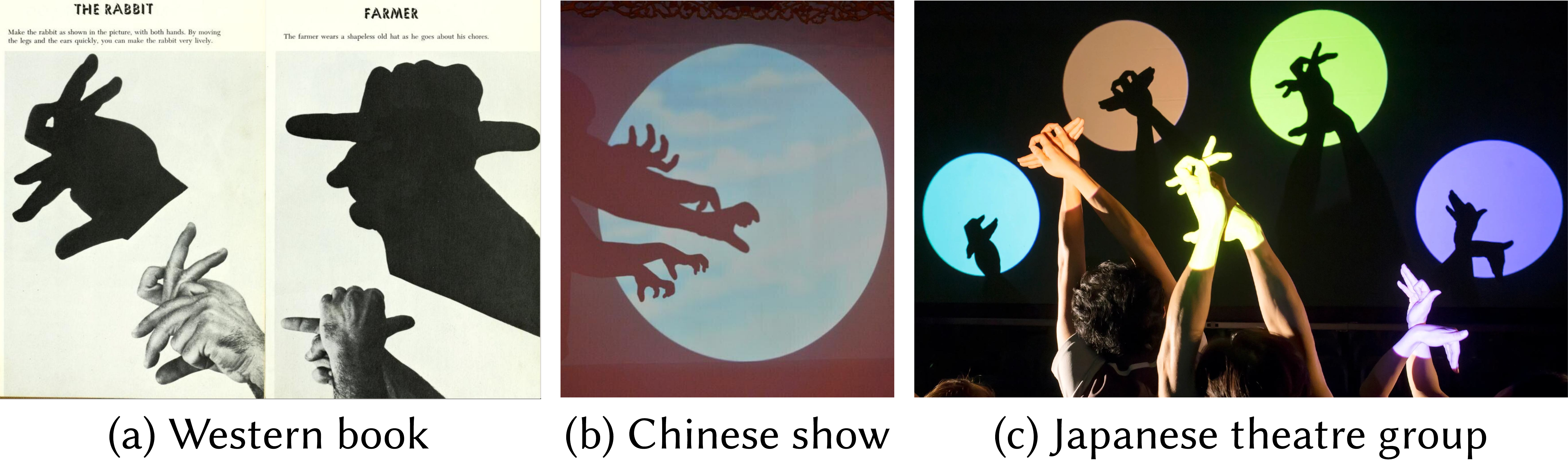}
    \caption{Hand-shadow examples from (a) the book ``The art of hand shadows''~\cite{albert1970art}, (b) traditional shadow play~\cite{shen2024shadow}, and (c) theatre group~\cite{kakashiza1952shadow}, from both the east and west.
    }
    \Description[]{}
    \label{fig:shadow_samples}
\end{figure}

Hand shadow art, also known as shadowgraphy~\cite{nikola1913complete}, is a captivating art form, in which the shadows cast by hands on a wall creatively reveal the shapes of various kinds of objects.
This art has a long and rich history across many cultures, since ancient times~\cite{albert1970art, frank1996fun}.
Its appeal lies in its simplicity, flexibility, and creativity.
With only a few easy-to-obtain items (\ie, hands, a light source, and a projective surface), one can create a wide variety of shadow shapes that mimic lifelike animals, plants, portraits, \etc; see some classic examples in Figure~\ref{fig:shadow_samples}.

\begin{figure}[t]
    \includegraphics[width=\linewidth]{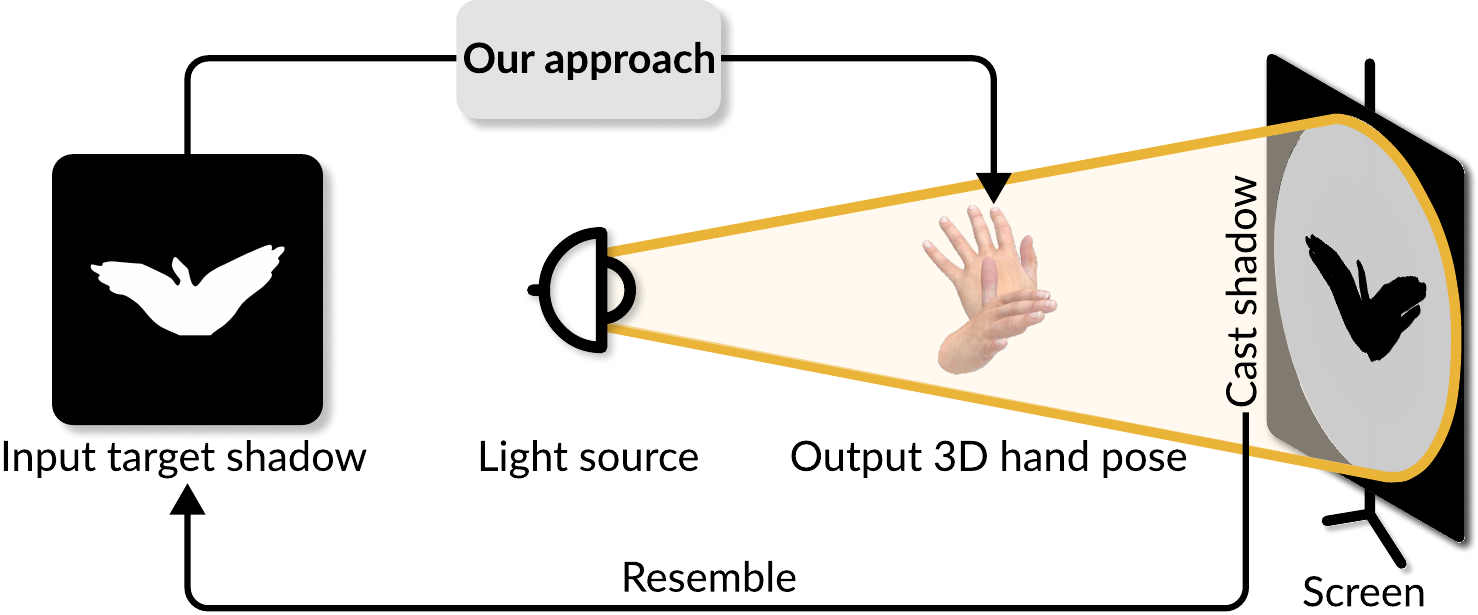}
    \caption{Illustrating our task. 
    Given a target shadow as the input, we aim to estimate the 3D poses of both the left and right hands, such that the two hands together can cast a shadow that closely resembles the input.
    Note that the light source and screen are fixed in the setup.}
    \Description[]{}
    \label{fig:goal}
\end{figure}

We are interested in solving an inverse problem; see also Figure~\ref{fig:goal}.
We develop a learning-based approach to find plausible bimanual hand poses that can cast shadows that closely resemble a given hand shadow mask.
Our approach enables users to explore a wide range of hand shadow forms, including animal-type shadows and extending to alphanumeric characters and even more intricate shapes, as shown in Figure~\ref{fig:teaser} for some of our results.

Finding bimanual hand poses for reproducing a target hand shadow is nontrivial.
First, the process is inherently ambiguous: the design space of 3D hand poses is huge, as a single shadow can often be produced by multiple different hand poses.
Second, we need to attend to both the shape and crucial features in the input, but the absence of color and texture in shadows makes the hand shape recovery ill-posed (\ie, significant changes in hand poses may not lead to any shadow changes, resulting in large plateau regions during pose-optimization via differentiable rendering). 
The fine-grained preservation of shadow features with restricted hand anatomy poses an additional challenge.
Moreover, from a model learning perspective, the scarcity of domain-specific shadow datasets further complicates the method design. 
A detailed elaboration on the problem definition, setup, and challenges can be found in Section~\ref{sec:overview}.
%

To meet these challenges, we build on one key insight:
The inverse hand shadow art problem can be addressed through two sub-tasks. 
Given a hand shadow mask, by (i) resolving the \textit{anatomical constraints} of two hands, it becomes feasible to recover anatomically correct hand shapes and poses; and (ii) resolving the \textit{semantic constraints} of shadow allows the coarse hand poses to reproduce a shadow shape that preserves the features of the input.
The decoupling allows solving the problem using only generic data with admissible hand configurations.
%

Specifically, to locate two hands from a bimanual hand mask, we first need to identify plausible 2D shapes for the left and right hands. 
It is challenging due to unknown overlapping regions and the mirror symmetry of the left and right hands.
Deterministic methods like segmentation are suboptimal, as they cannot account for diverse possible hand configurations. 
We address this with a probabilistic generative model to produce diverse but reasonable hand assignments.
Second, although shadows lack colors and textures, they provide shape priors.
To generalize single-hand pose recovery to the shadow-mask domain, we fine-tune an RGB-based hand pose recovery model in a semi-supervised manner, leveraging existing knowledge while addressing the absence of 3D annotations.
Last, to ensure that the reconstructed poses respect the most salient shadow features while maintaining anatomical plausibility, the optimization should prioritize key areas over the pixel-perfect alignment.

Based on these technical motivations, we design Hand-Shadow Poser, a three-stage framework to decouple the hand semantics from the anatomical constraints (imposed by the hands) and semantic constraints (imposed by the shadow shape):
(i)~A generative hand assignment module to predict plausible left-right hand shapes from the ambiguous shadow, by exploring diverse hypotheses via a conditional generative model.
(ii)~A generalized hand-shadow alignment module to robustly recover 3D poses of each hand-shape hypothesis to coarsely align with the shadow, followed by a similarity-driven strategy for selecting high-quality candidates.
(iii)~A shadow-feature-aware refinement module to iteratively optimize hand poses to reproduce salient features of shadow shape and ensure physical feasibility through carefully-designed constraints.
Our feed-forward models in the first two stages are trained on generic public hand datasets with a rich set of augmentation operations, freeing us from creating extensive specialized hand-shadow data for training.
%

To evaluate our approach, we built a benchmark, containing diverse 2D masks of varying complexity, including shadow arts from books, alphanumeric characters, and everyday objects from the MPEG-7 dataset~\cite{sikora2001mpeg}. 
We also define a comprehensive set of metrics to assess the quality of the reproduced shadows, focusing on perception, semantics, and salient characteristics. 
Quantitative comparisons with baselines, qualitative results, and user studies consistently exhibit the effectiveness and robustness of our approach.

Overall, our main contributions are as follows:
\begin{itemize}
    \item We introduce a comprehensive framework to compute hand shadow arts, covering a rich variety of shadow shapes.
    \item We design a three-stage pipeline to decouple the anatomical constraints imposed by the hand and the semantic constraints imposed by the shadow shape, enabling training merely on richly augmented generic public hand datasets.
    \item We formulate three novel components in our pipeline: generative hand assignment, generalized hand-shadow alignment, and shadow-feature-aware refinement.
    \item We construct a benchmark for evaluation, encompassing 210 shadow art forms with a variety of shapes, and introduce shadow-specific metrics for quality assessment. The effectiveness of our approach is demonstrated through extensive experiments, including quantitative evaluations, qualitative comparisons, and detailed user studies.
\end{itemize}

\noindent
\emph{The code and benchmark data of Hand-Shadow Poser will be publicly available at \url{https://github.com/hxwork/HandShadowPoser}.}
\section{Related Work}

\paragraph{Computational visual art}
Visual arts embrace a wide variety of genres, media, and styles, demanding profound human aesthetics and expertise in creations~\cite{wang2024diffusion}. A growing research has enabled computational generalization of various forms of visual arts, both 2D and 3D.
For example, the generation of stylized artworks such as 2D paintings~\cite{kopf2011depixelizing, chiu2015tone, binninger2024sd}, 3D scenes~\cite{zhang2022arf, haque2023instruct, liu2024stylegaussian}, 3D sculptures~\cite{liu2017image, yang2021wireroom}, and reliefs~\cite{schuller2014appearance}. 

In shadowgraphy, shadows projected onto the wall present expressive shapes and figures (\eg, animals), making it hard to believe that the shadow objects come merely from two human hands.
This line of art differs from general visual arts.
It is a specific genre characterized by a visual percept that differs from reality, such as optical illusion design~\cite{visualillusions}.
Specifically, our task lies at the intersection between 2D and 3D optical illusions.

\paragraph{2D optical illusion}
Numerous computational methods have been proposed to synthesize illusional images.
Oliva~\etal~\shortcite{oliva2006hybrid}, in an early attempt, generate hybrid images that exhibit appearance changes at different viewing distances; 
Chi~\etal~\shortcite{chi2008self} propose to arrange repeated asymmetric patterns to stimulate illusory motion perception.
Multiple efforts~\cite{chu2010camouflage, zhang2020deep, zhao2024lake} study camouflage, in which the goal is to compute images with certain imagery patterns subtly embedded in the images.
\cite{geng2024visual, burgert2024diffusion} explore multi-view illusion images whose appearance changes upon flips, rotations, skews, or jigsaw rearrangements.
Recently, Geng~\etal~\shortcite{geng2025factorized} generalize this technique to color saturation, motion blur, and inverse problems.

\paragraph{3D optical illusion}
Computational generation of 3D optical illusions can be roughly divided into two categories.
The first focuses on the digital fabrication of local microfacets for pattern display,~\eg, requiring certain lighting conditions.
Various mediums have been exploited such as spatially-varying reflectance functions~\cite{matusik2009printing}, 3D height fields~\cite{weyrich2009fabricating, wu2022computational}, microstructural stripe patterns~\cite{sakurai2018fabricating}, cellular mirrors~\cite{hosseini2020portal}, scratches on metal~\cite{shen2023scratch}, and refractive lenses~\cite{papas2012magic, zeng2021lenticular}.
Recently, researchers~\cite{perroni2023constructing, zhu2024computational} exploit self-occlusion to achieve view-dependent appearances without relying on an external light source. 

Our work is more related to the second category, which aims to generate a 3D shape that produces different forms of visual illusion.
Gal~\etal~\shortcite{collage3d2007} abstract input models into expressive 3D compound shapes with elements from a database.
Leveraging depth misperception caused by projection, Wu~\etal~\shortcite{wu2010modeling} create topological structures that seem impossible to exist, whereas Sugihara~\shortcite{sugihara2014design} creates solid shapes with slopes that appear to disobey the laws of gravity when a ball is placed on them.
Tong~\etal~\shortcite{tong2013mona} study the hollow-face illusion, in which a gradual deformation can be observed when walking around the object. 
Alexa and Matusik~\shortcite{alexa2010reliefs} study reliefs that approximate given images under certain illumination; 
Chandra~\etal~\shortcite{chandra2022designing} design a differentiable probabilistic programming language to create multiple illusions, including human faces that appear to change expressions under different lighting.
Creating 3D shapes with varying appearances from different view directions is initially explored in~\cite{sela2007generation}, which relies on geometric deformation from two input 3D models.
Keiren~\etal~\shortcite{keiren2009constructability} provide a theoretical analysis of the problem of constructing a triplet from a given set of three letters.
Intriguing variants are further studied,~\eg, in 3D shadow volumes~\cite{mitra2009shadow}, 3D crystals~\cite{hirayama2019projection}, and 3D wire sculptures~\cite{hsiao2018multi, qu2024wired, tojo2024fabricable}.

\begin{figure*}[t]
    \includegraphics[width=\linewidth]{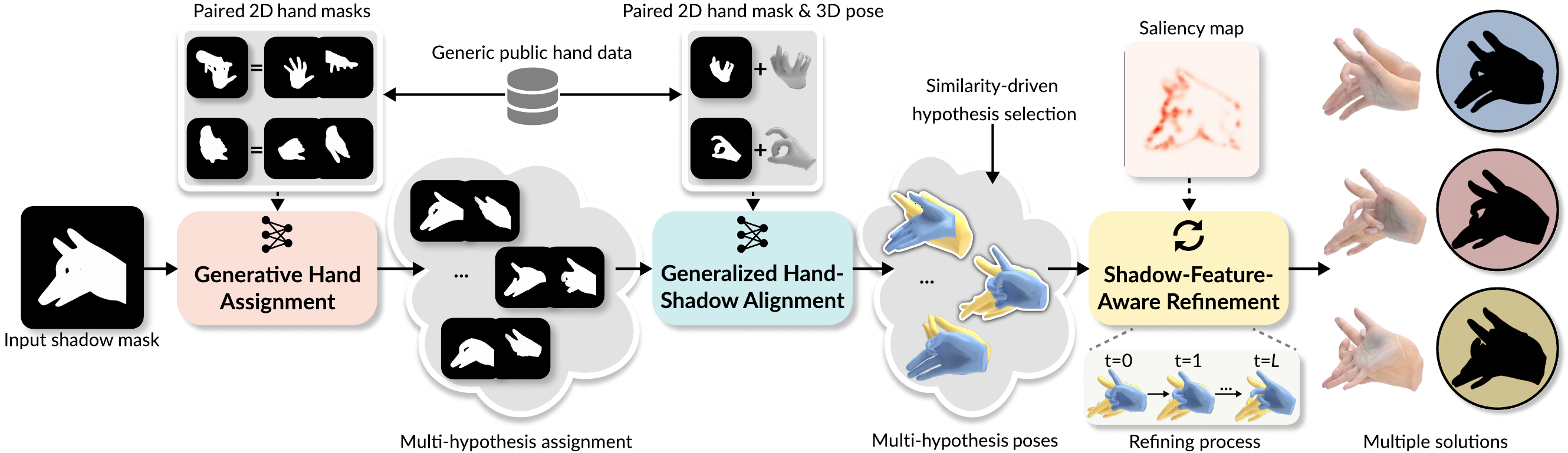}
    \vspace*{-6mm}
    \caption{Overview of our Hand-Shadow Poser, which consists of three key stages: (i) generative hand assignment, (ii) generalized hand-shadow alignment with similarity-driven hypothesis selection, and (iii) shadow-feature-aware refinement.
    }
    \Description[]{}
    \label{fig:pipeline}
\end{figure*}

\paragraph{Shadow art}
Shadows are the results of the interplay between light and objects.
Shadow art has been extensively explored to create expressive and illusional designs. 
Pellacini~\etal~\shortcite{pellacini2002user} present an interface for transforming shadows based on user requirements, whereas Mattausch~\etal~\shortcite{mattausch2013freeform} manipulate rendered shadows and apply the edited results in varying scene configurations.

Mitra~\etal~\shortcite{mitra2009shadow} design an algorithm to construct a 3D volume constrained by orthogonal shadow images as inputs, such that lighting the same solid from different specific directions interestingly creates different shadow patterns.
With a similar goal, Zhang~\etal~\shortcite{zhang20173d} develop a method to create 3D shadow art sculptures using a collection of real items. 
Chen~\etal~\shortcite{ballistic2017} propose a framework for generating animated target shadows using objects under ballistic motion.
Sadekar~\etal~\shortcite{Sadekar_2022_WACV} revisit shadow art with a differentiable rendering-based optimization.
Wang~\etal~\shortcite{wang2024neuralshadowart} further expand its potential and flexibility with implicit representations and joint optimization of lighting directions and screen orientations.
A special form, namely creating 3D wire sculptures based on multi-view sketches, is first explored by Hsiao~\etal~\shortcite{hsiao2018multi}.
Recently, Qu~\etal~\shortcite{qu2024wired} utilize flexible drawing capabilities from modern generative models, whereas Tojo~\etal~\shortcite{tojo2024fabricable} promote the fabricability of the reconstructed wires for 3D printing and support richer input controls. 
Gangopadhyay~\etal~\shortcite{knotgangopadhyay2024} deform a topological embedding of the circle in 3D space to create single- or multiple-view target shadows.

Instead of projecting from 3D shapes, some works create shadow art with manufactured planar-like devices.
Alexa and Matusik~\shortcite{alexa2012irregular} design a planar surface with holes to create self-shadows that induce single input images.
Bermano~\etal~\shortcite{bermano2012shadowpix} exploit walls and chamfers within a diffuse surface for producing self-shadowing effects that display multiple images under different views and lights.
Some variants study the casting of a single shadow onto an external plane to match various desired images.
Baran~\etal~\shortcite{baran2012manufacturing} present a multi-layer attenuator that casts different shadows depending on the light configuration.
To project and form a target pixel art image, Yue~\etal~\shortcite{yue2012pixel} arrange transparent sticks within a container to refract light; 
Zhao~\etal~\shortcite{zhao2016printed} design 3D-printed perforated lampshades to project continuous grayscale images, whereas Min~\etal~\shortcite{min2017soft} arrange multiple occluder layers to create a soft boundary shadow.

In this work, we aim to create prescribed shadows using human hands, inspired by the traditional hand shadow arts~\cite{nikola1913complete, albert1970art, frank1996fun}.
One of the most relevant works is~\cite{won2016shadow}, which generates human characters given 2D silhouette images by employing a nonlinear optimization to minimize the visual difference between the resultant and target shadow contours.
However, they require hints specified by professional actors to match specific points on the target contour with associated body parts, which is difficult to obtain for various shadow inputs in real-world scenarios.
Besides, directly applying a method following~\cite{won2016shadow} in our task tends to yield unsatisfactory results, since the optimization is inherently sensitive to initialization and easily converges to a local optimum.

Another closely related work is a short paper~\cite{gangopadhyay2023hand}, which solves a similar task to ours. 
They use differentiable rendering and directly minimize the image loss between the input and target shape.
Result-wise, it showcases only a few hand shadow examples.
Due to its optimization-based nature, similar issues are observed as in~\cite{won2016shadow}. 
Beyond the above two works, we present a novel and generalizable approach capable of (i)~covering a richer variety of hand shadow cases, (ii)~capturing salient characteristics of the target shadow, and (iii)~generatively proposing diverse hand poses with anatomical constraints. 
Also, we take optimization through differentiable rendering as a baseline in our comparison and show that differentiable rendering alone cannot achieve the results of our approach; see Section~\ref{sec:results_and_experiments} for the comparison experiment.
To our best knowledge, this is the first work that comprehensively studies the creation of hand shadow arts.

\paragraph{3D hand pose estimation from silhouettes}
Another closely-related research topic is 3D hand pose estimation from monocular RGB images~\cite{iqbal2018hand, zhang2019end, moon2020i2l, zhou2020monocular, zhang2021hand, chen2022mobrecon, xu2023h2onet, huang2023neural, zhou2024simple, pavlakos2024reconstructing}, a longstanding research task due to its significance in downstream applications.
Yet, very few attempts have been made to recover 3D hand poses from sparse 2D information, such as anatomical landmarks~\cite{ramakrishna2012reconstructing}, hand-drawn stick figures~\cite{lin2012sketching}, or binary masks~\cite{agarwal20043d, dibra2017human} that are conducted on human bodies.

\cite{lee2019silhouette} is the first work that estimates 3D single-hand pose from binary silhouettes, which requires additional depth supervision during the training stage. 
Under the same setting, Chang~\etal~\shortcite{chang2023mask2hand} achieve comparable performance as state-of-the-art RGB-based and depth-based methods without relying on depth information.
However, both works focus on single-hand inputs. Directly applying their method to bimanual hand masks remains challenging since we need to estimate the locations of the two hands in the input while the input is simply a binary mask, in which the hand shapes are obscured.
Thus, we should not only solve the ill-posed problem of locating a pair of non-intersecting interacting hands within a single mask, but also collectively estimate the poses of the two hands to reproduce the target shadow.
\section{Overview}
\label{sec:overview}

\paragraph{Problem definition}
Figure~\ref{fig:goal} illustrates our task.
The input is a target shadow represented as a binary mask, whereas the outputs are the 3D poses of the left and right hands represented by the MANO~\cite{romero2022embodied} hand model. 
With a light source and screen plane, we aim to inversely find the 3D poses of the hands positioned between them, such that the projected hand shadow on the screen can closely match the given target shadow. 

We further clarify the setup. 
Creating hand shadows with clear and sharp boundaries requires a small, intense light source and a flat projection screen, as outlined in classical references~\cite{nikola1913complete, albert1970art}. 
The light source and screen remain fixed, while the hands are adjusted in between. 
The hands, light source, and screen are horizontally and vertically aligned to minimize distortion.
For simplicity, we focus on scenarios with only two hands, without considering other body parts and additional object items.

\paragraph{Challenges}
To achieve our goal, one straightforward approach is to directly optimize the hand poses by minimizing the visual difference between the cast shadow and the target shadow~\cite{won2016shadow, gangopadhyay2023hand}. 
However, there exist several key challenges outlined below:
\begin{enumerate}[label=(\roman*),leftmargin=*]
    \item \textit{Initialization sensitivity}: Optimization-based methods are sensitive to initialization and prone to converge to local optima. 
    Providing a good initial condition is a crucial step towards a successful result and fast optimization~\cite{finn2017model}, yet it remains challenging due to the huge search space.
    \item \textit{Feature preservation}: It is infeasible to match every pixel of the projected and the input shadows due to limited hand anatomy. 
    Instead, the most distinctive features of the input mask should be retained, whereas manually specifying hints~\cite{won2016shadow} is impractical. 
\end{enumerate}
Another straightforward approach is to train a neural network model to predict the interacting hand poses from the target shadow in a feed-forward manner, utilizing prior distributions learned from training datasets.
Yet, this process is also nontrivial due to the following challenges:
\begin{enumerate}[label=(\roman*),leftmargin=*]
    \item[(iii)] \textit{Dataset scarcity}: Given the scarcity of annotated hand shadow art datasets, the model must be robust and generalizable, without relying on prior knowledge from a specific data domain, to avoid labor-intensive data preparation.
    \item[(iv)] \textit{Results diversity and robustness}: The same given shadow could be produced by multiple different hand poses.
    Especially when two (left and right) hands are considered, there can be many different choices.
    Identifying diverse yet reasonable results introduces another challenge to the network design.
\end{enumerate}
\paragraph{Overview of our Hand-Shadow Poser}
Figure~\ref{fig:pipeline} gives an overview of our approach, which has the following three stages:
(i) the \textit{generative hand assignment} stage assigns diverse reasonable left-right 2D hand shapes (masks) to cover different parts of the shadow in the input binary mask (Section~\ref{sec:hand_assignment});
(ii) the \textit{generalized hand-shadow alignment} stage recovers a coarse 3D hand pose of each single-hand binary mask and automatically selects the high-quality ones for the subsequent stage (Section~\ref{sec:pose_estimation}); and 
(iii) the \textit{shadow-feature-aware refinement} stage iteratively refines the coarse 3D hand poses to make their shadows resemble the input, considering physical plausibility (Section~\ref{sec:finger_alignment}). 
In the end, we take our approach to work on diverse hand shadow examples from our benchmark, and conduct a series of evaluations to demonstrate the quality of our results and the effectiveness of the proposed designs.
\section{Generative Hand Assignment}
\label{sec:hand_assignment}
The first stage aims to find rough 2D hand shapes (masks) with reasonable anatomy to match the input shadow.
We name this task \emph{hand assignment},~\ie, to assign each hand to cover different parts of the target shadow.
In particular, we do not require the 3D hand poses for shadow matching in this stage.
Here, the main challenges are due to the lack of information in the input, which is just a binary mask, and also to the many different possible hand shapes that may eventually match and form the target shadow.
\begin{figure}[t]
    \includegraphics[width=\linewidth]{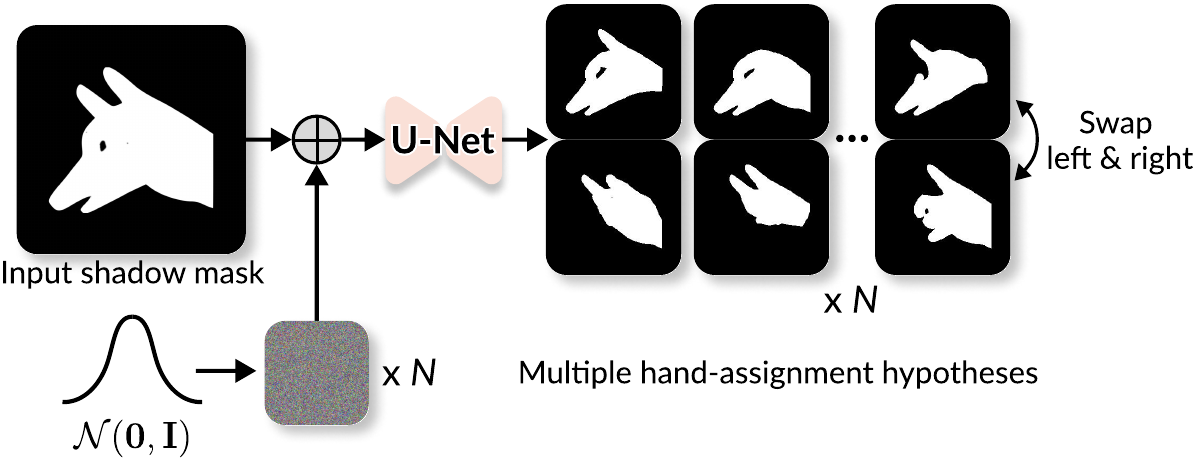}
    \vspace{-5mm}
    \caption{Our generative approach for hand assignment introduces diversity for exploring more different 2D hand shapes.
    Likewise, we additionally swap the left and right hands to model mirror symmetry in hand assignment.
    }
    \Description[]{}
    \label{fig:stage1_model}
\end{figure}
\paragraph{An initial attempt}
At the beginning of this research, we tried an image segmentation approach,~\ie, to classify each pixel in the input shadow mask as the left hand, right hand, or both (overlapping). 
Specifically, we adopted the network architecture in~\cite{liu2023efficientvit} and trained it on a mixture of datasets with rendered segmentation labels.
Then, we observed several drawbacks.
First, due to the deterministic nature, using a segmentation model discourages capturing the uncertainty in shadow-to-hand mapping.
Second, segmentation emphasizes pixel-level accuracy, so the trained model tends to focus excessively on the hand shapes rather than exploring their global cues.
Last, the absence of color and texture in the input largely raises the complexity of network learning compared to conventional segmentation tasks.
Hence, this approach leads to inferior performance, as shown later in Section~\ref{sec:results_and_experiments}.
\paragraph{Our generative approach}
To overcome the above issues, we propose to formulate a generative approach for hand assignment.
By doing so, we aim to introduce diversity in the results to address the ambiguity in shadow-to-hand mapping; see Figure~\ref{fig:stage1_model}.
Also, we aim to make the network learning easy, so that the network model can better attend to the overall hand shapes than pixel-level recovery.

Method-wise, we design the generative hand assignment model based on the conditional denoising diffusion probabilistic model~\cite{ho2020denoising}.
To learn the reverse diffusion process, we adopt the classifier-free guidance~\cite{ho2022classifier} for shadow-controlled multi-hypothesis generation.
By concatenating the input binary mask $\hat{\mathbf{M}}$ with the intermediate noisy output $\mathbf{x}_t$ at timestamp $t$ (ranging from 0 to $T$), our network model can progressively reach the final assignment $\mathbf{x}_0 \!\in\! \mathbb{R}^{H \times W \times 2}$ (\ie, a two-channel image with height $H$ and width $W$) using the denoising model $f_{\text{assign}}(\cdot)$:
\begin{equation}
    \mathbf{x}_0=f_{\text{assign}}(\hat{\mathbf{M}},\operatorname{PE}(t)) 
\end{equation}
where $t$ is encoded through positional embedding (PE)~\cite{NIPS2017_3f5ee243}.
The assigned left- and right-hand masks $\mathbf{M}_l$ and $\mathbf{M}_r$ are then obtained from $\mathbf{x}_0$ using
\begin{equation}
    \mathbf{M}_{l}, \mathbf{M}_{r} = \operatorname{Split}(\mathbf{x}_0),
\end{equation}
where $\operatorname{Split}(\cdot)$ denotes the channel split operation.

In addition, to speed up the inference, we employ DDIM~\cite{song2020denoising} to sample $\mathbf{x}_t$ at arbitrary timestamps.
At inference, $N$ initial noise vectors $\{\mathbf{x}_T^i| i \in {1,...,N} \}$ are randomly sampled from the Gaussian distribution $\mathcal{N}(\mathbf{0}, \mathbf{I})$ to produce diverse hand assignment results.
Also, we swap the left and right hands (Figure~\ref{fig:stage1_model}) to further enrich the diversity.
Specifically, a U-Net is adopted as the denoising model $f_{\text{assign}}(\cdot)$, in which we employ an encoder with four downsample blocks, each with two residual blocks; an attention mechanism; a downsampling layer; and a decoder with four upsample blocks in a structure similar to the encoder.  Further, a residual block is employed to yield the two-channel map $\mathbf{x}_0$.

Furthermore, following~\cite{karras2022elucidating}, we employ the L2 distance between the predicted and ground-truth values as the training loss, in which we separately calculate the left and right masks:
\begin{equation}
\mathcal{L}_{\text{assign}}=\frac{\bar{\alpha}_t}{1-\bar{\alpha}_t} \sum_{*\in\{l,r\}}||\mathbf{M}_*-\hat{\mathbf{M}}_*||_2^2,
\end{equation}
where $\bar{\alpha}_t$ denotes the total noise variance at step $t$, as defined in~\cite{song2020denoising}.
\begin{figure}[t]
    \includegraphics[width=\linewidth]{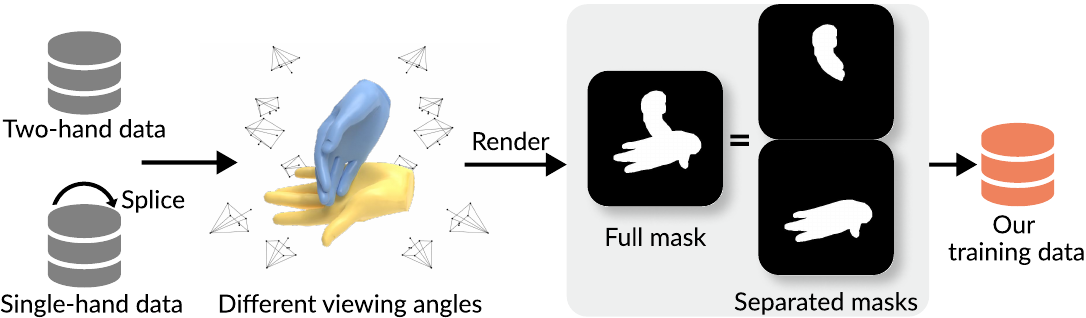}
    \vspace{-5mm}
    \caption{
    To prepare training data for generative hand assignment, we augment existing two-hand and single-hand datasets by (i) randomly splicing left- and right-hand samples in single-hand datasets to synthesize more two-hand samples, and (ii) rendering two-hand samples in different views.
    }
    \Description[]{}
    \label{fig:stage1_data}
\end{figure}

\paragraph{Training data}
Benefiting from our decoupled pipeline design, the hand assignment model mainly needs to learn the knowledge of 2D hand shapes instead of shadow semantics. 
Hence, to prepare for the training data, we propose to leverage the rich 2D and 3D ground-truth labels in existing public hand datasets~\cite{moon2020interhand2, zuo2023reconstructing,zhang2017hand,zimmermann2019freihand} that were built for various other purposes~\eg, hand pose estimation and tracking.
By doing so, we can train our model using generic hand datasets, including also synthetic ones~\cite{li2023renderih}.

To do so, we augment existing hand datasets to provide the supervision for network model training in two aspects, as illustrated in Figure~\ref{fig:stage1_data}.
First, since two-hand datasets are scarce, compared with single-hand ones, we randomly splice (combine) left- and right-hand samples from single-hand datasets, following~\cite{zuo2023reconstructing}, thereby synthesizing more diverse interacting poses of varying levels of hand overlap, which could occur in real hand shadow art scenarios.
Second, we render the 3D interacting hand meshes from multiple perspectives to enrich the diversity of viewing angles.
By these means, we can substantially increase both the quantity and diversity of two-hand samples for network training.
\section{Generalized Hand-Shadow Alignment}
\label{sec:pose_estimation}

Given the estimated left- and right-hand masks $\mathbf{M}_l$ and $\mathbf{M}_r$, the second stage aims to construct the 3D poses (\ie, the 61 MANO coefficients that represent the hand orientation, axis-angle 3D poses of 15 hand joints, hand shape, and 3D coordinate of the wrist joint) of the left and right hands $(\theta_l, \beta_l, t_l)$ and $(\theta_r, \beta_r, t_r)$, such that the resulting hand poses provide a coarse 3D hand alignment with the target shadow.
Importantly, we do not require capturing the fine-grained shadow features at this stage.
Rather, we need coarse 3D predictions from rough 2D hand shapes of diverse poses.

\begin{figure}[t]
    \includegraphics[width=\linewidth]{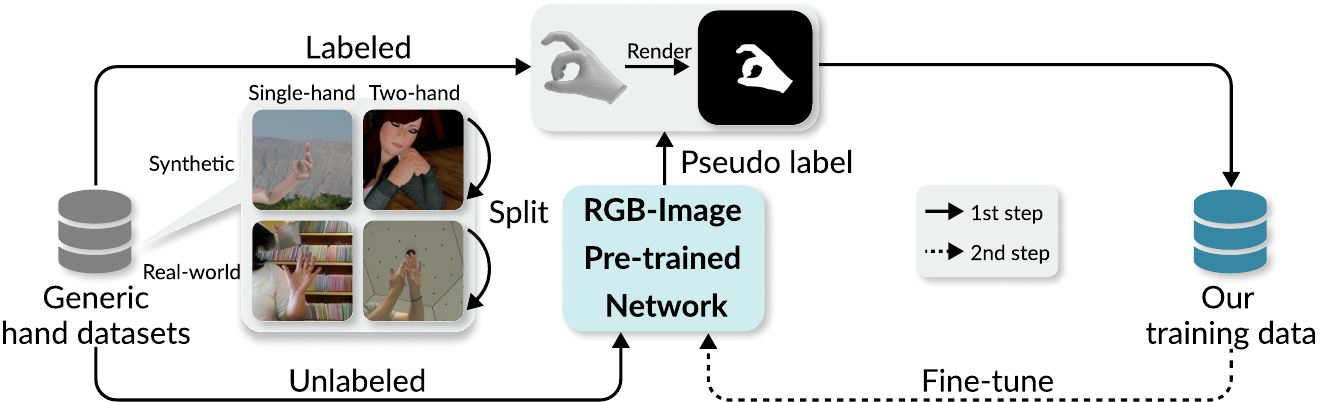}
    \caption{To prepare training data for generalized hand-shadow alignment, we propose to use a semi-supervised learning strategy as illustrated above, considering the use of both labeled and unlabeled hand samples from both real and synthetic datasets.
    With this strategy, we can produce a dataset with paired 2D masks and 3D poses for fine-tuning our network.
    }
    \Description[]{}
    \label{fig:stage2_data}
\end{figure}

Considering that single-hand poses are relatively easier to infer than collectively estimating interacting hand poses, we propose to narrow down the search space by predicting the pose of each hand separately. 
Here, we train the pose recovery network $f_{\text{align}}(\cdot)$ to predict the MANO representation of each hand from its mask:
\begin{equation}
    \theta_*, \beta_*, t_* = f_{\text{align}}(\mathbf{M}_*),
\end{equation}
where subscript $*$ denotes l (left) or r (right).
Yet, to make a good prediction is still nontrivial for two reasons.
First, it is hard to recover 3D hand poses from the colorless masks, providing only sparse information. 
Second, the input hand mask is simply a rough approximation of the actual hand shape; due to shadow ambiguity, we need a robust network model to overcome the uncertainty.

\paragraph{A generalized approach}
To meet these challenges, we aim for a generalizable and robust performance from two perspectives: (i) generalizing well-learned knowledge from the RGB-image domain to the shadow mask domain; and (ii) leveraging large data priors in existing data to generalize and handle rough 2D hand shapes.
The above considerations motivate us to adopt a large-scale fully transformer-based design~\cite{dosovitskiy2020image}.
First, to fully generalize knowledge from the RGB image domain, we initialize the network model $f_{\text{align}}$ with the pre-trained weights from~\cite{pavlakos2024reconstructing} to leverage vast data prior learned from extensive RGB image data.
To effectively handle the uncertainty in the inputs, we further fine-tune the network using a comparable magnitude of binary mask images from a large collection of generic hand datasets, including both real and synthetic data with single and interacting hands.
Specifically, we adopt the Vision Transformer (ViT)~\cite{dosovitskiy2020image} as the network backbone, which takes embeddings of image patches as input. The output tokens are then fed into a transformer decoder to regress the MANO parameters by cross-attending to a single query token. 
Last, the hand mesh and its relative translation to the camera can be converted through a MANO layer. 

\paragraph{Model training}
We adopt loss functions similar to~\cite{dosovitskiy2020image} to supervise the network training:
\begin{equation}
\begin{aligned}
    & \mathcal{L}_{\text{align}}^{\text{3D}} = \|\theta-\hat{\theta}\|_2^2 + \|\beta-\hat{\beta}\|_2^2 + \|\mathbf{J}^{\text{3D}}-\hat{\mathbf{J}}^{\text{3D}}\|_1 \\ 
    \mathrm{and} \
    & \mathcal{L}^{\text{2D}}_{\text{align}} = \|\mathbf{J}^{\text{2D}}-\hat{\mathbf{J}}^{\text{2D}}\|_1,
\end{aligned}
\end{equation}
where $\mathbf{J}^{\text{3D}}$ denotes the 3D joint coordinates converted from the predicted MANO parameters; 
$\mathbf{J}^{\text{2D}}$ denotes their projections onto the image space by using the camera intrinsics; and 
the quantities with the hat superscript $\hat{}$ are the ground-truth labels.
Here, $\mathcal{L}^{\text{2D}}_{\text{align}}$ is utilized to promote consistency in the output image space, following~\cite{dosovitskiy2020image}.

In the training process, we first split the left and right hands from the existing two-hand data to obtain more single-hand training samples. 
Since not all data samples are paired with MANO-represented ground truths, we take a semi-supervised learning strategy as illustrated in Figure~\ref{fig:stage2_data},~\ie, employing the RGB-image pre-trained network on unlabeled images to estimate the MANO coefficients as pseudo labels, then rendering the results to produce paired binary hand masks. 
With this approach, we can avoid the need for highly accurate pseudo labels associated with the original images, as our focus is on ensuring the anatomical correctness of the hand poses. 
Second, we take these samples, together with the labeled samples, to form our dataset for network training.
\paragraph{Similarity-driven hypothesis selection}
The 2D hand shape hypotheses from the previous stage are fed into the pose recovery network to obtain 3D hand poses,~\ie, pose hypotheses.
However, this process does not take into account the quality of the hypotheses, which may largely degrade the performance of the next stage.
Since ground truths are not available at inference, we thus formulate a similarity-driven strategy to evaluate and select pose hypotheses.

\begin{figure}[t]
    \includegraphics[width=\linewidth]{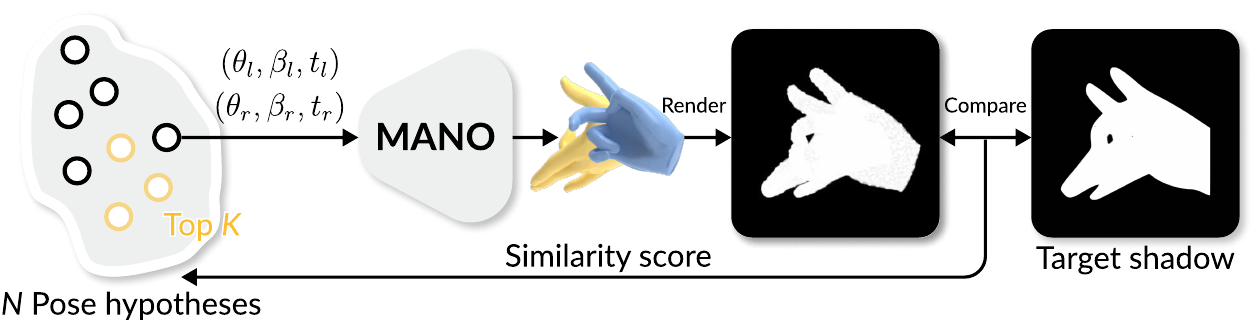}
    \caption{Our similarity-driven hypothesis selection strategy, taking a render-and-compare approach to favor pose hypotheses of the highest quality.
    }
    \Description[]{}
    \label{fig:stage2_selection}
\end{figure}

Overall, our idea is to maximize the similarity between the target shadow and the reproduced shadow, by a render-and-compare approach.
That is, we first project and render each pose hypothesis (\ie, its 3D hand mesh) into a binary mask, and then calculate its perceptual similarity to the input mask using LPIPS~\cite{zhang2018perceptual} and DINOv2~\cite{oquab2023dinov2} semantic-based scores. 
By sorting all $N$ hypotheses based on their similarity scores, we can then select the top $K$ hypotheses for refinement in the next stage; see Figure~\ref{fig:stage2_selection}.
Details about similarity scores are introduced in Section~\ref{sec:metrics}.
\section{Shadow-Feature-Aware Refinement}
\label{sec:finger_alignment}
\begin{figure}[t]
    \includegraphics[width=\linewidth]{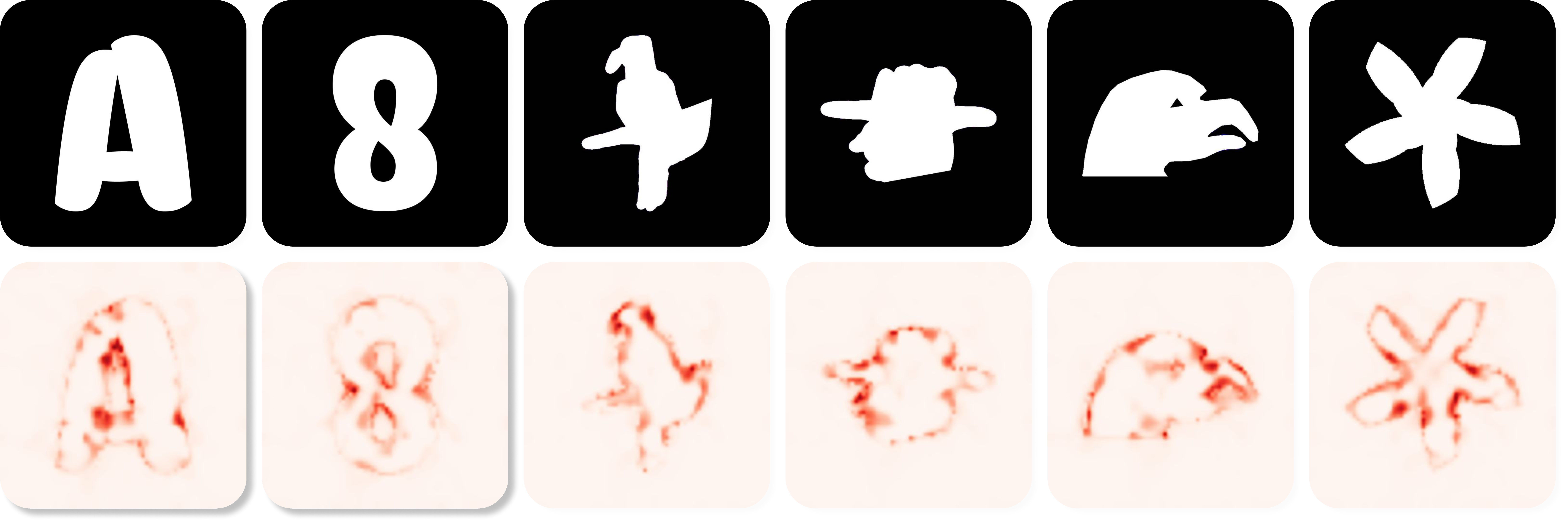}
    \caption{
    Top row: target shadows.
    Bottom row: extracted saliency maps, in which characteristic shadow features are highlighted in red.
    }
    \Description[]{}
    \label{fig:saliency}
\end{figure}

\begin{figure}[t]
    \includegraphics[width=\linewidth]{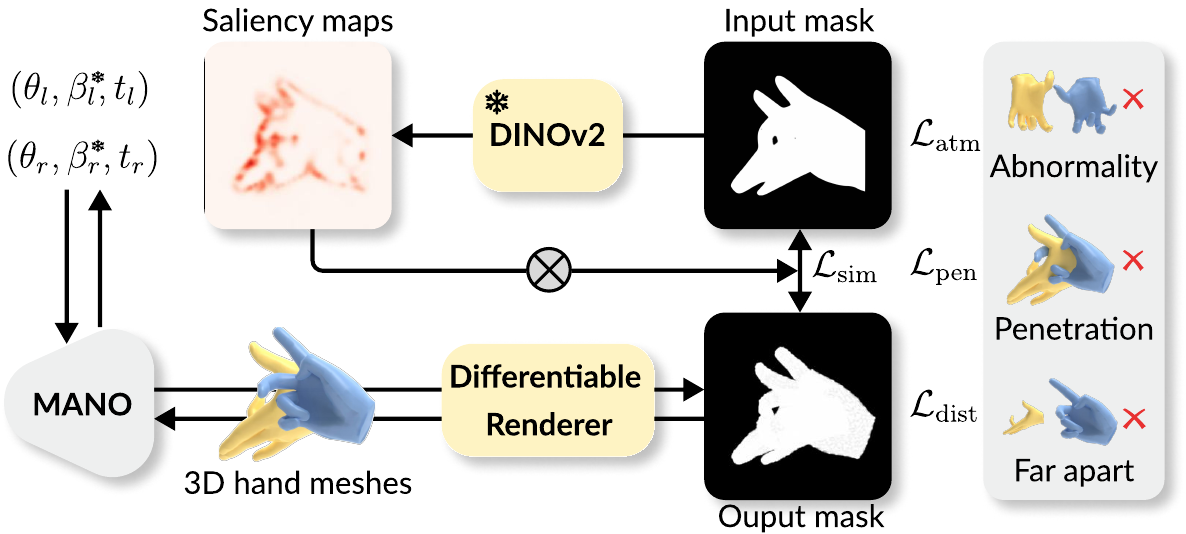}
    \caption{Our shadow-feature-aware refinement iteratively optimizes the 3D hand poses to align with features in the input using the DINOv2-based saliency guidance, while considering physical constraints in terms of anatomy, penetration, and hand-to-hand distance.
    }
    \Description[]{}
    \label{fig:stage3}
\end{figure}

To successfully reproduce the target shadow, the final 3D hand poses need to attend to the global shape of the shadow, as well as to the details or shadow features; see~\eg, the beak of the parrot and the eyes of the eagle and wolf in  Figures~\ref{fig:saliency} and~\ref{fig:stage3}.
Hence, the final stage aims to refine the coarse 3D hand poses to make their projections perceptually more similar to the target shadow, considering particularly the shadow features together with the anatomical constraints.

Method-wise, the overall approach is based on differentiable rendering.
That is, we first create a binary mask of the hands by projecting the coarse 3D hand meshes from the previous stage. 
Then, we iteratively optimize the joint angles and wrist positions of the two hands with their shape parameters fixed, mimicking real-world hand pose adjustments; see Figure~\ref{fig:stage3}. 
Importantly, beyond the differentiable rendering in~\cite{gangopadhyay2023hand}, where pose initialization is rarely considered and often leads to suboptimal results, the coarse outputs from 
our first two stages provide a good initial condition for the optimization process to achieve a faster and better convergence~\cite{finn2017model, rajeswaran2019meta, Lee2020Learning}.
This can be attributed to the prior knowledge of hands brought about by our decoupling design. 

Below, we introduce four carefully-crafted constraints for the optimization.
The first constraint aims to maximize the similarity between the input and rendered masks, with saliency guidance for preserving the shadow features.
To favor physically-plausible hand poses, we further incorporate the other three constraints, considering anatomy, penetration, and hand-to-hand distance.

\begin{figure*}[t]
    \includegraphics[width=\linewidth]{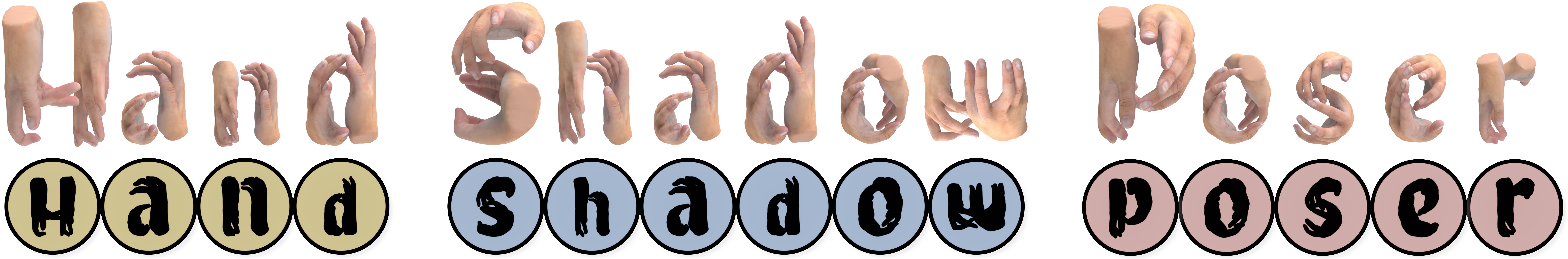}
    \vspace*{-5mm}
    \caption{A gallery of ``Hand Shadow Poser'' created by our Hand-Shadow Poser on various uppercase and lowercase letters.
    }
    \Description[]{}
    \label{fig:gallery_alphanumeric}
\end{figure*}

\paragraph{(i) Similarity constraint with saliency guidance}
In the optimization, our main goal is to minimize the misalignment between the rendered mask $\mathbf{M}$ and the input mask $\mathbf{\hat{M}}$.
Directly constraining their image discrepancy with L1 or L2 loss can lead to suboptimal results, due to the significant gap in flexibility between the limited range of hand joint movement and the expressive capacity of shadows. 
Rather than aligning every pixel equally, we prioritize preserving the shadow features.
Also, this process is desired to be automated, eliminating the need to manually specify the hint points in~\cite{won2016shadow}.
To this end, we propose to leverage DINOv2~\cite{oquab2023dinov2}, a powerful pre-trained vision model, to first locate prominent features in the input shadow shape; see again Figure~\ref{fig:saliency}).
Given the input mask $\hat{\mathbf{M}}$, we assign varying levels of importance to different shadow regions based on the extracted saliency map:
\begin{equation}
    \mathcal{L}_{\text{sim}} = \sum \left(1 + \operatorname{DINO}(\hat{\mathbf{M}})\right) \odot \left|\mathbf{M} - \hat{\mathbf{M}}\right|,
\end{equation}
where $\operatorname{DINO}(\cdot)$ represents DINOv2's attention heatmap extraction and $\odot$ is the Hadamard product.

\paragraph{(ii) Anatomy constraint}
To mitigate a pose's abnormality, we adopt the twist-splay-bend frame in~\cite{yang2021cpf} by projecting the rotation axis to three independent axes, then computing the penalization of the abnormal axial components on each joint as $\mathcal{L}_{\text{atm}}$.
Please refer to~\cite{yang2021cpf} for the details.

\paragraph{(iii) Penetration constraint}
Inspired by~\cite{jiang2021hand}, we identify vertices of one hand that are inside the other hand (the set is denoted as $\mathbf{P}_{\text{in}}$) and define the inter-penetration loss as their distances to the closest vertices on the other hand:
\begin{equation}
    \mathcal{L}_{\text{inter-pen}} = 
    \frac{1}{|\mathbf{P}_{\text{in}}|}
    \sum_{p\in \mathbf{P}_{\text{in}}}min_{i}\|p-\mathbf{V}_{i}\|_2^2,
\end{equation}
where $\{\mathbf{V}_i\}$ represents mesh vertices of the hand being penetrated.
For self-penetration, we adopt the conic distance fields approximation of meshes in~\cite{tzionas2016} to penalize the depth of intrusion, denoted as $\mathcal{L}_{\text{self-pen}}$.
The final penetration loss $\mathcal{L}_{\text{pen}}$ is a sum of $\mathcal{L}_{\text{inter-pen}}$ and $\mathcal{L}_{\text{self-pen}}$.
For the detailed calculation of $\mathcal{L}_{\text{self-pen}}$, please refer to~\cite{ballan2012motion, tzionas2016}.

\paragraph{(iv) Hand-to-hand distance constraint}
With the above constraints, we optimize the validity of two hand poses and achieve the desired shape in the projection space. 
However, since the optimization is not sensitive to movements along the depth axis after the projection, the resulting hand meshes can become too far apart along the depth axis relative to the light source.
Moreover, even a relatively moderate distance,~\eg, one meter, can significantly complicate the process of creating the hand shadows. 

Concerning this, we propose a new loss term to constrain the distance between the wrist joints of the two hands. 
Empirically, we penalize this distance when it exceeds a certain threshold $\tau_{\text{dist}}$.
\begin{equation}
\mathcal{L}_{\text{dist}} =
    \begin{cases}
     \|t_l - t_r\|_2^2 & \text{if }\|t_l - t_r\|_2^2 \geq \tau_{\text{dist}} \\
    0 & \text{otherwise},
    \end{cases}
\end{equation}
where $t_{l}$ and $t_{r}$ are the 3D joint coordinates of the left-hand and right-hand wrists, respectively.

\begin{figure*}[t]
    \includegraphics[width=\linewidth]{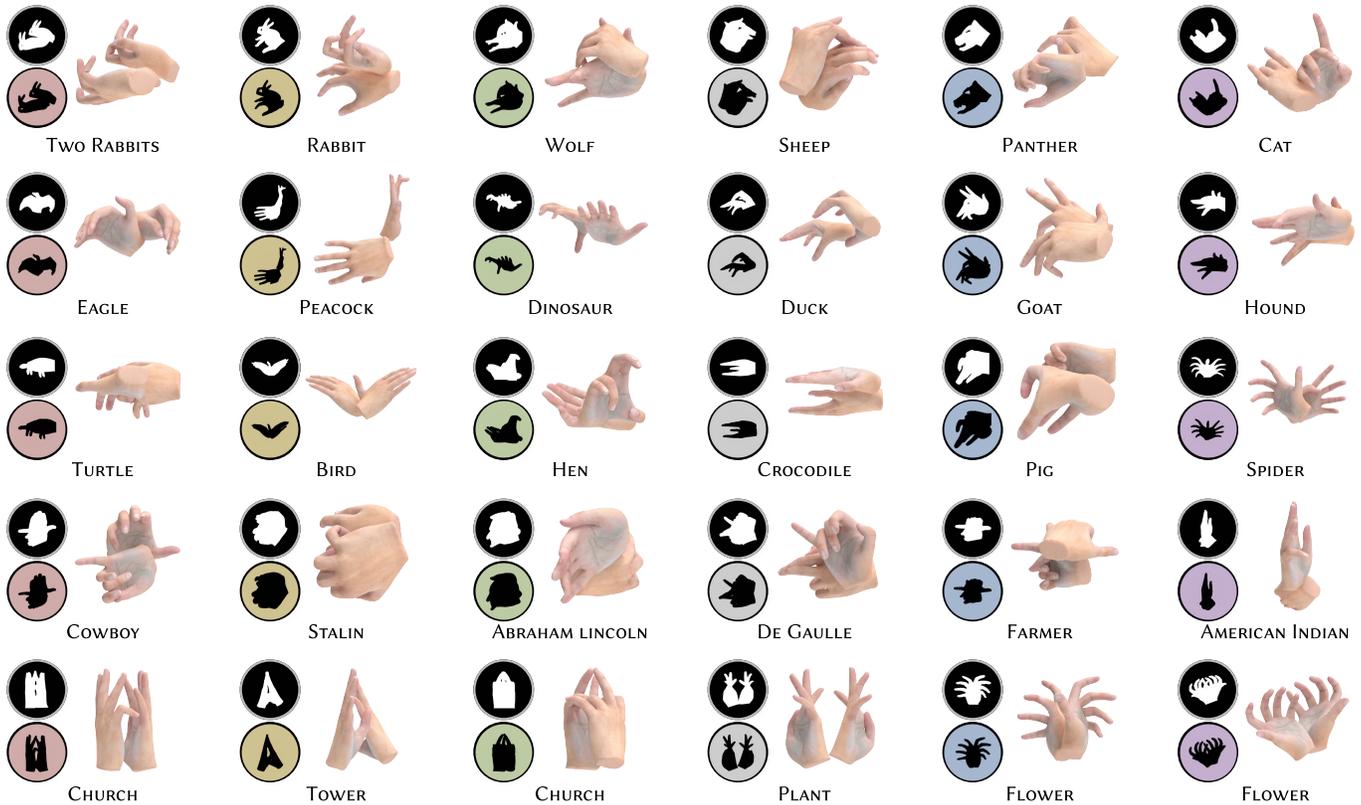}
    \vspace*{-5mm}
    \caption{A gallery showcasing the results of our Hand-Shadow Poser on real hand-shadow-art shapes (C2), which are obtained from the following books~\cite{albert1970art, frank1996fun, nikola1913complete}, covering a wide range of shapes encompassing animals, human portraits, buildings, and plants.
    For each case: the top left shows the target shadow, the bottom left shows our reproduced shadow, whereas the right shows our produced 3D hand poses.}
    \Description[]{}
    \label{fig:gallery_realshadow}
\end{figure*}
\paragraph{Optimization} 
The final objective is a weighted sum of the terms:
\begin{equation}
\begin{aligned}
    \operatorname{min}_{\theta_l, t_l, \theta_r, t_r} & [w_{\text{sim}} \mathcal{L}_{\text{sim}} + w_{\text{atm}}\mathcal{L}_{\text{atm}}+
     w_{\text{pen}}\mathcal{L}_{\text{pen}} + w_{\text{dist}}\mathcal{L}_{\text{dist}}].
\end{aligned}
\end{equation}
where $w_{\text{sim}}$, $w_{\text{atm}}$, $w_{\text{pen}}$, and $w_{\text{dist}}$ are hyperparameters. 
Further, we adopt Adam~\cite{2015-kingma} for gradient-descent-based optimization, which ends after $L$ iterations. 
\section{Results and Experiments}
\label{sec:results_and_experiments}

\subsection{Experimental Setup}
\label{sec:experimental_setup}

\paragraph{Baselines}
We compare our Hand-Shadow Poser with three baselines:
\begin{itemize}
    \item \emph{Baseline 1} optimizes the 3D hand poses with differentiable rendering as in~\cite{gangopadhyay2023hand}, with random initialization three times, and then picks the best one based on the similarity metrics. 
    \item \emph{Baseline 2} uses a single neural network to directly regress the coarse pose of the interacting hands from the input shadow mask, followed by the same optimization as \emph{Baseline 1}.
    \item \emph{Baseline 3} replaces the generative model in Stage 1 with a segmentation model, as described in Section~\ref{sec:hand_assignment}.
\end{itemize}

\paragraph{Training datasets}
We prepare the training data for the feed-forward models from multiple public hand datasets, including single- and two-hand datasets.
Specifically, to train the generative hand assignment model in Stage 1 (Section~\ref{sec:hand_assignment}), we prepare pairs of two-hand masks and left-right hand masks from InterHand2.6M~\cite{moon2020interhand2}, RenderIH~\cite{li2023renderih}, and Two-hand 500K~\cite{zuo2023reconstructing}. 
We also follow~\cite{zuo2023reconstructing} to randomly combine single-hand data in~\cite{gomez2019large, moon2020interhand2, zhang2017hand, zimmermann2017learning, zimmermann2019freihand}. 
Leveraging the multi-perspective augmentation strategy, we obtain 7.7M data samples in total for generative model training.

\begin{figure*}[t]
    \includegraphics[width=\linewidth]{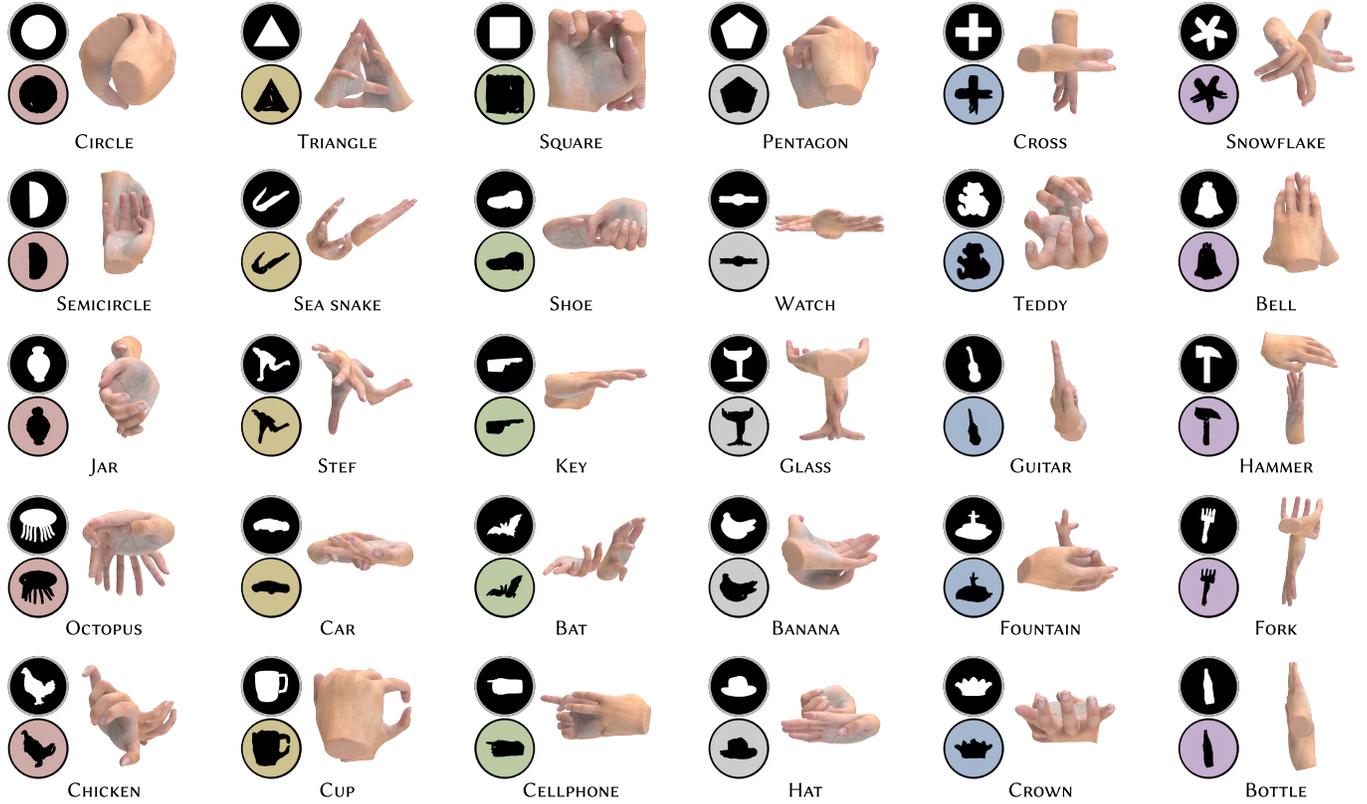}
    \vspace*{-5mm}
    \caption{A gallery of hand shadow arts created by our Hand-Shadow Poser for shapes of diverse everyday objects (C3) from~\cite{sikora2001mpeg} and the Internet. 
    For each case, the top left shows the target shadow, the bottom left shows our reproduced shadow, whereas the right shows our produced 3D hand poses.}
    \Description[]{}
    \label{fig:gallery_arbitrary}
\end{figure*}
\begin{figure}[t]
    \includegraphics[width=0.99\linewidth]{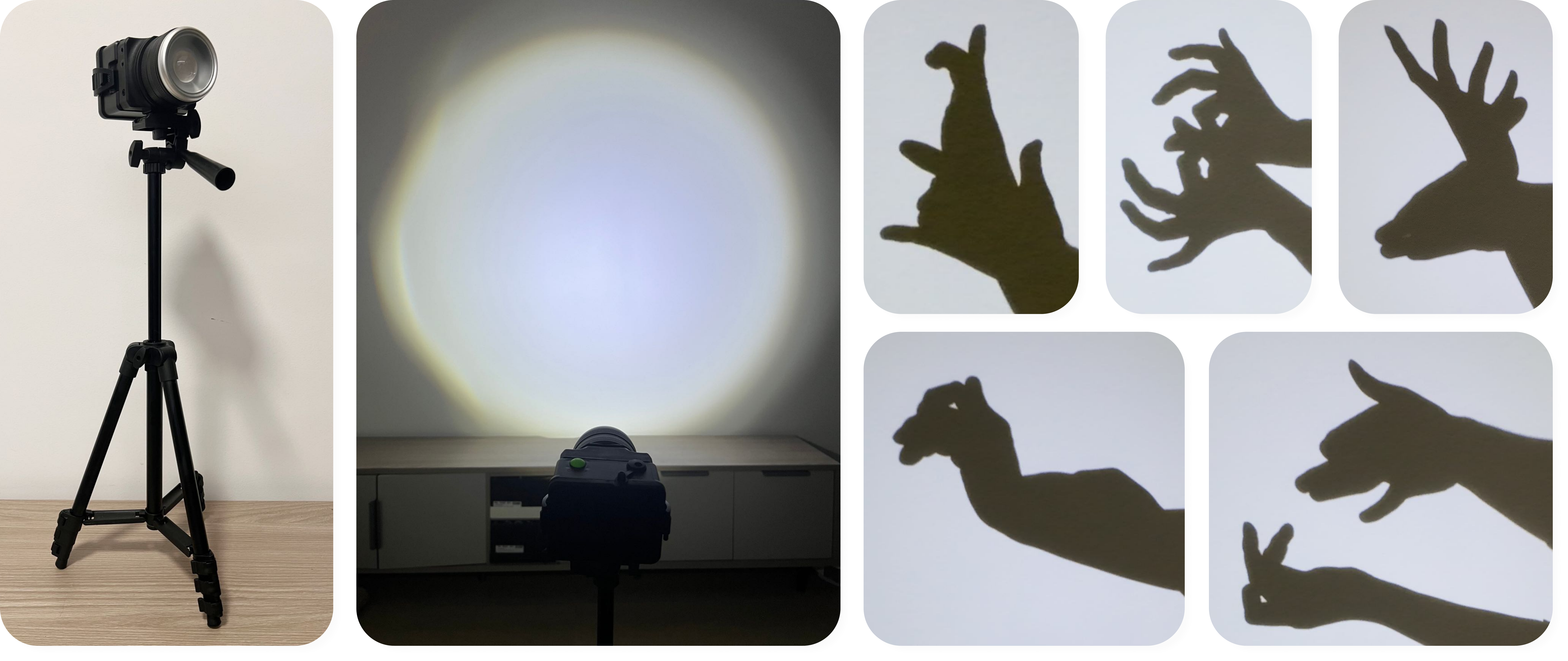}
    \caption{
    Left: our physical setup.
    Right: real shadows created for \textsc{Kangaroo}, \textsc{Crab}, \textsc{Deer}, \textsc{Camel}, and \textsc{Fox chases rabbit}.
    }
    \Description[]{}
    \label{fig:real_shadow}
\end{figure}
\begin{figure}[t]
    \includegraphics[width=\linewidth]{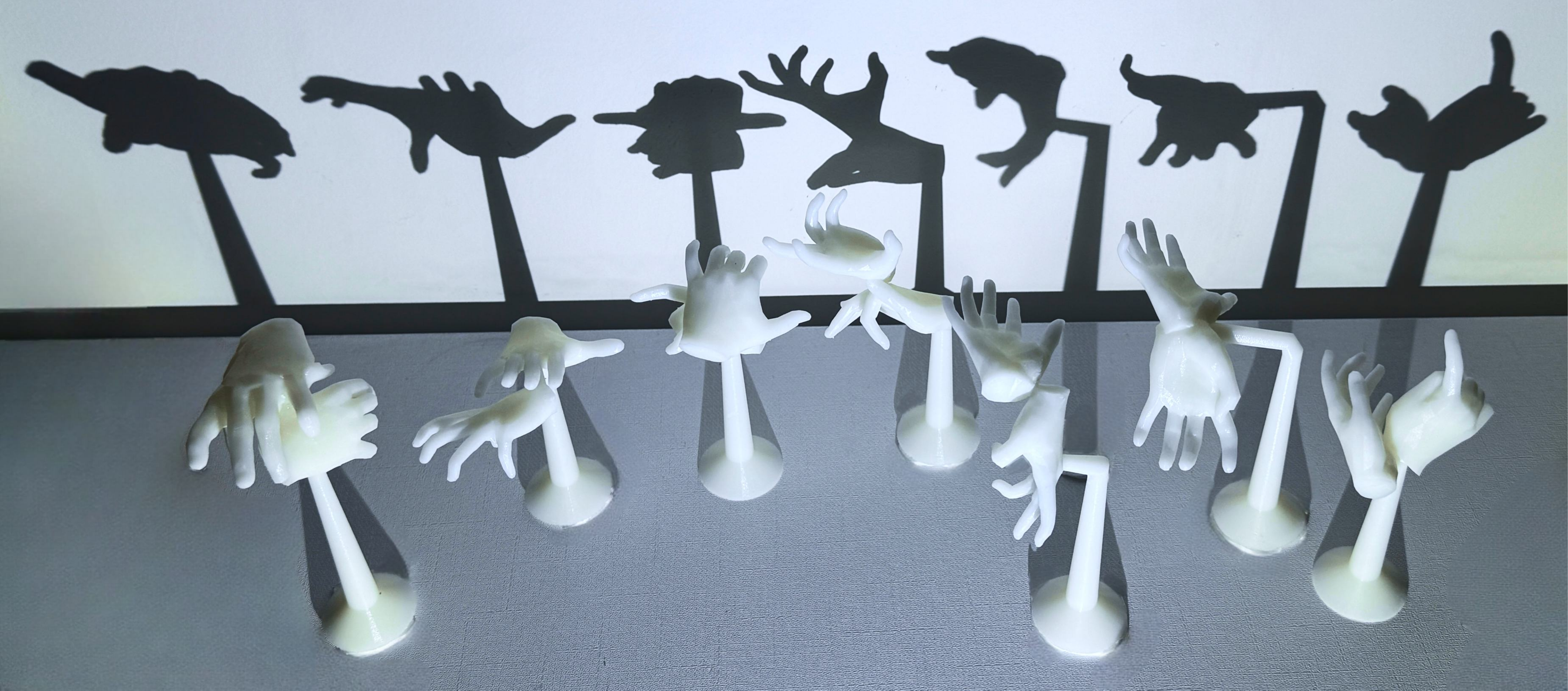}
     \caption{ 
     3D-printing seven of our 3D hand pose results, which are reconstructed for reproducing the following shadow shapes: \textsc{Tortoise}, \textsc{Dinosaur}, \textsc{Farmer}, \textsc{Deer}, \textsc{Dolphin}, \textsc{Elephant}, and \textsc{Cat} (left to right).
     }
    \Description[]{}
    \label{fig:3d_printing}
\end{figure}

To train the large-scale transformer network in Stage 2 (Section~\ref{sec:pose_estimation}), we use a large collection of public datasets, following~\cite{pavlakos2024reconstructing}, FreiHAND~\cite{zimmermann2019freihand}, HO3D~\cite{hampali2020honnotate}, MTC~\cite{xiang2019monocular}, RHD~\cite{zimmermann2017learning}, InterHand2.6M~\cite{moon2020interhand2}, H2O3D~\cite{hampali2020honnotate}, DexYCB~\cite{chao2021dexycb}, COCO WholeBody~\cite{jin2020whole}, Halpe~\cite{fang2022alphapose}, and MPII NZSL~\cite{simon2017hand}. 
We additionally incorporate RenderIH~\cite{li2023renderih} and Two-hand 500k~\cite{zuo2023reconstructing} by splitting them into left- and right-hand data. 
The final training set consists of 2.7M samples.

\paragraph{Evaluation benchmark}
\label{sec:benchmark}
We constructed a benchmark of 210 binary mask images, covering a wide variety of hand shadow shapes, for quantitative and qualitative evaluation.
The dataset includes 62 alphanumeric characters (C1), 87 real hand-shadow-art shapes (C2) from~\cite{nikola1913complete, albert1970art, frank1996fun}, 
and 61 shapes of diverse everyday objects (C3) from~\cite{sikora2001mpeg} and Internet.
For more examples, please refer to our supplementary material.
To the best of our knowledge, this is the first work that collects such a diverse and challenging set of hand shadow shapes for a systematic analysis of computing hand-shadow arts.

\paragraph{Metrics}
\label{sec:metrics}
On the other hand, we propose to use the following five metrics to evaluate the visual similarity between the generated shadow and the input shadow:
(i)~\emph{LPIPS}: We adopt LPIPS~\cite{zhang2018perceptual} to measure the perceptual similarity based on the deep features from AlexNet~\cite{krizhevsky2012imagenet}. 
Building on CLIP~\cite{radford2021learning}, we employ two other similarity metrics, including
(ii)~\emph{CLIP-Global}, which evaluates the image-level semantic similarity by employing the CLIP image encoder to map the shadow mask image to the CLIP space and then calculating the cosine distance; 
and (iii)~\emph{CLIP-Semantic}, which computes the cosine similarity between the CLIP text embedding of the input shadow's class description (\eg, ``rabbit'') and the image embedding of the generated shadow, to assess its level of alignment with the text semantics.
Considering the model's sensitivity to text, we adopt the officially-released CLIP code by scaling the original similarities by a factor of 100, followed by a softmax operation to obtain the logit scores for the reproduced hand shadow mask of each baseline and our method, respectively.
(iv)~\emph{DINO-Global}: Similar to~\emph{CLIP-Global}, we leverage DINOv2~\cite{oquab2023dinov2} for feature extraction to evaluate the visual similarity at a global scale.
(v)~\emph{DINO-Semantic}: Additionally, to remedy the ignorance of the above metrics to local characteristics, we design this metric to measure the preservation of local shadow features between the output mask $\mathbf{M}$ and input mask $\hat{\mathbf{M}}$:
\begin{equation}
    \textit{DINO-Semantic} = \frac{\sum(\mathbf{M}-\hat{\mathbf{M}})\odot \mathds{1} (\operatorname{DINO}(\mathbf{\hat{M}})> \tau_{\text{semantic}})}{\sum\mathds{1} (\operatorname{DINO}(\mathbf{\hat{M}})> \tau_{\text{semantic})}},
\end{equation}
where $\tau_{\text{semantic}}$ is set to 0.1 by default and $\mathds{1} (\cdot)$ is the indicator function.
These metrics together provide a comprehensive evaluation of the quality of the reproduced shadows.
\paragraph{Implementation details}
We adopt Blender~\cite{blender} to create all the hand-shadow-art scenes. 
For shadow projection, we set up a spotlight with a beam radius of 0.001 and an angle of 15$\degree$, with 1000 W power. The distance from the light source to the projection plane is set to 2.5 m. 
The focal length of the perspective camera in differentiable rendering is set to 1 m, aligning with the one used in~\cite{pavlakos2024reconstructing}.
To avoid projection deviation, the camera is positioned at the same world coordinates as the light source, facing the same direction towards the projective surface.
We implemented our method using PyTorch~\cite{paszke2019pytorch} and adopted the Adam optimizer for both training the feed-forward models (Stages 1 and 2) and optimizing the hand orientations, poses, and translations (Stage 3). All experiments were conducted on eight NVIDIA Tesla V100 GPUs. 

Specifically, for Stage 1, we use a batch size of 48 with a learning rate of 1e-4 and train the network model for 20 epochs. 
The input images are resized to $256\!\times\!256$, with a random rotation in $[0, 360\degree]$ and scaling in $[0.75, 1.25]$ for online data augmentation.
During the inference, the number of reverse steps for DDIM is 1,000.

In Stage 2, the model is fine-tuned for 10 epochs using a batch size of 8 and a learning rate of 1e-5. 
The input images are resized to $256$$\times$$256$ with the online augmentation strategies in~\cite{pavlakos2024reconstructing}.
For similarity-driven hypotheses selection, we set the default number of candidate hypotheses $N$ to 20 and the number of selected poses $K$ to 3.
The training processes for the first two stages take 3 days and 1 day, respectively.
\begin{table*}[t]
    \centering
    \caption{Quantitative comparison between baselines and our Hand-Shadow Poser.
    C1: Alphanumeric characters; C2: Real hand-shadow-art shapes; and C3: Shapes of everyday objects. The first five columns present metrics (\emph{LPIPs},~\emph{CLIP-Global},~\emph{CLIP-Semantic},~\emph{DINO-Global}, and~\emph{DINO-Semantic}) used to evaluate the visual similarity, while the last two columns (\emph{Human-Global} and~\emph{Human-Semantic}) are human-related metrics from the user study.
    }
    \resizebox{\textwidth}{!}{
        \setlength\tabcolsep{3pt}
        \begin{tabular}{l|cc|cc|cc|cc|cc|cc|cc}
        \toprule
        \multirow{2}{*}{Methods} & \multicolumn{2}{c|}{\emph{LPIPS}
        $\downarrow$} & \multicolumn{2}{c|}{\emph{CLIP-Global} $\uparrow$} & \multicolumn{2}{c|}{\emph{CLIP-Semantic}
        $\uparrow$} & \multicolumn{2}{c|}{\emph{DINO-Global} $\uparrow$} & \multicolumn{2}{c|}{\emph{DINO-Semantic}  $\downarrow$} & \multicolumn{2}{c|}{\emph{Human-Global} $\uparrow$} & \multicolumn{2}{c}{\emph{Human-Semantic} $\uparrow$} \\
        \cmidrule{2-15}
        & C1 / C2 / C3 & Avg. & C1 / C2 / C3 & Avg. & C1 / C2 / C3 & Avg. & C1 / C2 / C3 & Avg. & C1 / C2 / C3 & Avg. & C1 / C2 / C3 & Avg. & C1 / C2 / C3 & Avg. \\
        \cmidrule{1-15}
        Baseline 1 & 0.19 / 0.20 / 0.17 & 0.19 & 0.84 / 0.91 / 0.88 & 0.88 & 0.11 / 0.29 / 0.15 & 0.20 & 0.51 / 0.61 / 0.51 & 0.55 & 0.66 / 0.81 / 0.64 & 0.72 & 2.27 / 2.11 / 2.42 & 2.27 & 2.35 / 2.34 / 2.55 & 2.41 \\ 
        \midrule
        Baseline 2 & 0.20 / 0.19 / 0.17 & 0.19 & 0.83 / 0.91 / 0.88 & 0.88 & 0.10 / 0.18 / 0.22 & 0.17 & 0.47 / 0.63 / 0.51 & 0.55 & 0.70 / 0.84 / 0.66 & 0.74 & 2.03 / 2.49 / 2.36 & 2.29 & 2.15 / 2.42 / 2.42 & 2.33 \\
        \midrule
        Baseline 3 & 0.18 / 0.17 / 0.15 & 0.16 & 0.89 / 0.93 / 0.91 & 0.91 & 0.36 / 0.20 / 0.27 & 0.27 & 0.65 / 0.74 / 0.65 & 0.69 & 0.61 / 0.75 / 0.59 & 0.66 & 3.66 / 3.07 / 3.53 & 3.42 & 3.49 / 3.07 / 3.45 & 3.33 \\ 
        \midrule
        \textbf{Ours} & \textbf{0.15} / \textbf{0.15} / \textbf{0.13} & \textbf{0.14} & \textbf{0.91} / \textbf{0.95} / \textbf{0.93} & \textbf{0.93} & \textbf{0.42} / \textbf{0.33} / \textbf{0.35} & \textbf{0.36} & \textbf{0.71} / \textbf{0.80} / \textbf{0.75} & \textbf{0.78} & \textbf{0.47} / \textbf{0.67} / \textbf{0.49} & \textbf{0.56} & \textbf{4.66} / \textbf{4.45} / \textbf{4.47} & \textbf{4.53} & \textbf{4.55} / \textbf{4.21} / \textbf{4.29} & \textbf{4.35} \\
        \bottomrule
        \end{tabular}}
    \label{tab:quantitative_results}
\end{table*}

\begin{figure}[t]
    \includegraphics[width=\linewidth]{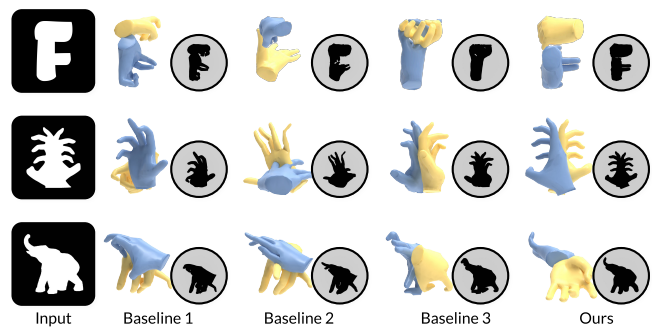}
    \caption{Comparing hand-shadow-art results produced by the three baselines and by our Hand-Shadow Poser.}
    \Description[]{}
    \label{fig:qualitative_comparison}
\end{figure}

For Stage 3, $w_{\text{sim}}$, $w_{\text{atm}}$, $w_{\text{pen}}$, and $w_{\text{dist}}$ are empirically set to 10.0, 1.0, 1.0, and 1.0, respectively, whereas $\tau_\text{dist}$ is set to 0.5 by default.
A Gaussian blur with a kernel size of $15\!\times\!15$ is applied to the extracted saliency map.
We optimize the hand parameters with a learning rate of 1e-3 and decay it by 0.5 at the 3,000th iteration, with the maximum number of iterations $L$ set to 6,000.

\subsection{Evaluation}
\label{sec:quantitative}
\paragraph{Gallery}
We present our visual results for three classes of shapes: alphanumeric characters (C1) in Figure~\ref{fig:gallery_alphanumeric}, real hand-shadow-art shapes (C2) in Figure~\ref{fig:gallery_realshadow}, and shapes of diverse everyday objects (C3) in Figure~\ref{fig:gallery_arbitrary}, which exhibit varying levels of complexity.
Details about the three classes of shapes are presented in Section~\ref{sec:benchmark}.
These results showcase the remarkable versatility of our method in reproducing many different kinds of object shapes, covering animals, plants, human portraits, logos, daily-used tools, numbers, letters,~\etc. 
More results are provided in the supplementary material.
\paragraph{Human demonstration}
\label{para:human_demo}
Next, we present some real shadow results.
Figure~\ref{fig:real_shadow} shows our physical setup and some example real shadows produced by human hands using a spotlight, demonstrating the feasibility of our method in practical scenarios.

\paragraph{3D fabrication} 
Further, we 3D-printed several 3D hand-poses results at a scale of 1:4 relative to the size of normal human hands.
In detail, we printed a horizontal tube invisible from the front view to join the two disconnected hands and another vertical/L-shaped tube from the bottom to the middle of the horizontal tube to support the two-hand sculpture.
Figure~\ref{fig:3d_printing} shows the results.
Interestingly, these 3D-printed hands look like simple hands in the real world, but if we look at them from a specific angle or shine a light in this direction, we can observe the shapes hidden by the hand sculptures.

\paragraph{Qualitative comparison}
\label{para:quality_comparison}
In Figure~\ref{fig:qualitative_comparison}, we visually compare our method with the baselines on three target shadows: \textsc{F}, \textsc{Flower}, and \textsc{Elephant}. 
Our method can create high-fidelity results that adhere to the input shadow details, highlighted by successfully preserving the key characteristics of shape, including all the petals in \textsc{Flower} and the bending nose of the \textsc{Elephant}. 
Though \emph{Baseline 3} can provide a relatively better initialization than the other baselines (based on the relative hand positions in \textsc{Flower}), it still struggles to produce a well-aligned shadow, primarily due to the segmentation model's ignorance of learning hand shapes from a global view.

\begin{figure}[t]
    \includegraphics[width=\linewidth]{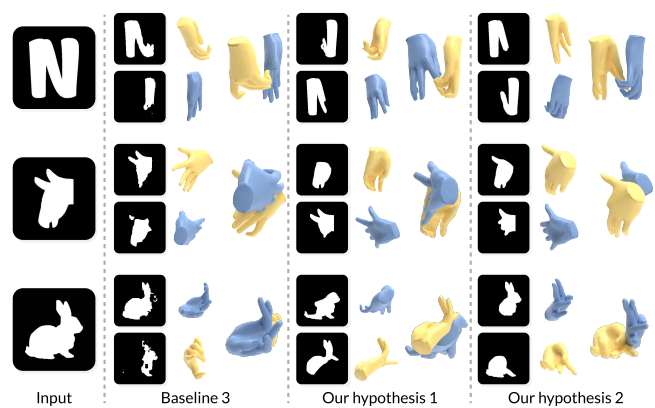}
    \caption{Qualitative comparison between Hand-Shadow Poser and \emph{Baseline 3}. 
    Hand assignment results (left), coarse 3D hand poses (middle), and refined 3D hand poses (right) are shown for each case.}
    \Description[]{}
    \label{fig:hand_assignment_comparison}
\end{figure}

\paragraph{Generative~\vs segmentation}
To further evaluate our generative approach against the segmentation-based approach, we compare the hand assignment results and recovered 3D hand poses before/after refinement from our method with those in \emph{Baseline 3}.
From Figure~\ref{fig:hand_assignment_comparison}, we can see that our method is capable of producing diverse hand shapes of higher quality, in terms of smoothness and completeness in the hand masks, 
whereas the segmentation model in \emph{Baseline 3} fails for complex input shapes like the \textsc{Stanford bunny} ( see the over-segmentation artifacts).
Consequently, under the same alignment and optimization process, our method yields multiple 3D hand poses with superior alignment to the target shadows, demonstrating the benefits of our generative formulation. 

\paragraph{Quantitative comparison}
\label{para:quantitative_comparison}
We compare the three baselines with our method on all three classes of shapes in the benchmark. 
Table~\ref{tab:quantitative_results} reports the full results on seven metrics, including the five metrics presented in Section~\ref{sec:metrics} and two human-related metrics to be presented later in the user study.
Our method achieves the best results for all metrics and classes, showing its superiority in preserving the shape and features in the target shadows for both perceptual and semantic similarity. 

\paragraph{Runtime comparison}
Our Hand-Shadow Poser includes two stages of feed-forward models and a test-time optimization; therefore, we report their running efficiency separately.
For fairness, all runtime measurements were taken on a single NVIDIA RTX 2080Ti GPU.

First, the models in the previous two stages have an average processing time of 3 minutes and 30 milliseconds per shape.

For Stage 3, as discussed in~\cite{hospedales2021meta}, initialization plays a crucial role in preventing being stuck at local minima during the optimization. 
To show the advantages of our initialization from feed-forward models, we report the number of iterations and runtime for the optimization to converge in each baseline and our method.
Specifically, the terminating condition is empirically set as (i) when the result's quality exceeds a certain~\emph{LPIPS} score calculated as the mean~\emph{LPIPS} of all four methods reported in Table~\ref{tab:quantitative_results}, or (ii) when the optimization reaches a maximum of 6,000 iterations.

\begin{table}[t]
    \centering
    \caption{Mean iteration number and runtime for optimization convergence.}
    \resizebox{\linewidth}{!}{
        \setlength\tabcolsep{10pt}
        \begin{tabular}{l|cc|cc}
        \toprule
        \multirow{2}{*}{Methods} & \multicolumn{2}{c|}{\emph{\# Iterations} $\downarrow$} & \multicolumn{2}{c}{\emph{Time (in seconds)} $\downarrow$} \\
        \cmidrule{2-5}
        & C1 / C2 / C3 & Avg. & C1 / C2 / C3 & Avg. \\
        \cmidrule{1-5}        
        Baseline 1 & 5809 / 5328 / 4792 & 5314 & 348 / 319 / 287 & 318 \\ 
        \midrule
        Baseline 2 & 5532 / 4883 / 3691 & 4728 & 331 / 293 / 221 & 283 \\ 
        \midrule
        Baseline 3 & 1845 / 1933 / 1705 & 1841 & 110 / 116 / 102 & 110 \\ 
        \midrule
        \textbf{Ours} & \textbf{1216} / \textbf{548} / \textbf{1234} & \textbf{945} & \textbf{73} / \textbf{32} / \textbf{74} & \textbf{56} \\ 
        \bottomrule
        \end{tabular}
        }
    \label{tab:runtime}
\end{table}
\begin{table}[t]
    \centering
    \caption{Ablation study on the shadow-feature-aware refinement module.}
    \resizebox{\linewidth}{!}{
        \setlength\tabcolsep{2pt}
        \begin{tabular}{l|ccccc}
        \toprule
        Methods & \emph{LPIPS} $\downarrow$ & \emph{CLIP-Global} $\uparrow$ & \emph{CLIP-Semantic} $\uparrow$ & \emph{DINO-Global} $\uparrow$ & \emph{DINO-Semantic} $\downarrow$  \\
        \cmidrule{1-6}
        w/o refinement & 0.19 & 0.92 & 0.30 & 0.69 & 0.76  \\ 
        \midrule
        w/o saliency & 0.15 & 0.93 & 0.36 & 0.74 & 0.59 \\ 
        \midrule
        \textbf{Full (ours)} & \textbf{0.14} & \textbf{0.94} & \textbf{0.40} & \textbf{0.78} & \textbf{0.56}  \\ 
        \bottomrule
        \end{tabular}
        }
    \label{tab:ablation}
\end{table}

In Table~\ref{tab:runtime}, we can observe a significantly reduced running time of our method, particularly around 1/6 compared to random initialization in \emph{Baseline 1}. 
Also, our method requires only half the optimization time of \emph{Baseline 3}, demonstrating the superiority of taking a generative approach to produce more effective initial configurations. 
Combined with the results in Section~\ref{para:quantitative_comparison}, it indicates that the coarse hand poses from our first two stages not only yield better-optimized hand shadows but also accelerate the convergence.
The efficacy of initialization from our method lies in the generalizability and robustness offered by the generative hand assignment and the generalized hand-shadow alignment module.

\subsection{User Studies}
We conducted two user studies to assess human preferences of the results produced by our approach versus the baselines, and also how our approach assists humans in creating hand shadow art.
\paragraph{Participants}
We invited 13 volunteers aged 22 to 30 with no professional experience in hand shadow art. The participants are divided into two groups. The first group has 10 participants (5 males and 5 females) who helped to assess the visual quality of the rendered shadows and to perform human demonstrations.
The other group (2 males and 1 female) served as judges in the second study to score the quality of human demonstrations after a brief tutorial session.

\paragraph{Metrics}
For quality comparison with the baselines, the rendered shadows are evaluated in two aspects: 
(i) global shape similarity, which measures how similar the reproduced shadows are compared with the target shadows (\emph{Human-Global}); and
(ii) local details similarity, which measures the details preservation (\emph{Human-Semantic}). 
The metrics are evaluated in a Likert scale from 1 (worst) to 5 (best). 
For the human demonstrations, we employ \emph{Human-Semantic} to measure the quality of the reproduced real shadows.
Also, we record the time taken by the participants in reproducing each shadow.

\begin{figure}[t]
    \includegraphics[width=\linewidth]{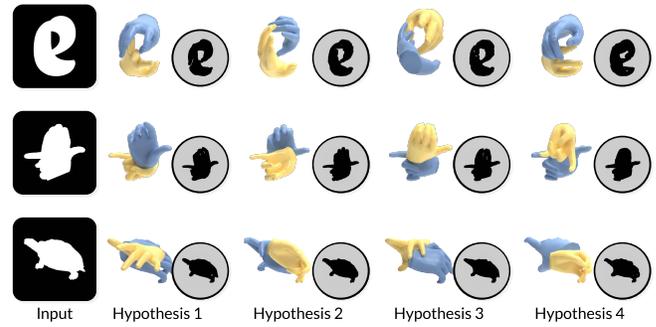}
    \caption{Result diversity brought by our Hand-Shadow Poser. 
    Note that the uniqueness is not limited to mirror symmetry, see results in the last row.}
    \Description[]{}
    \label{fig:diversity}
\end{figure}
\begin{figure*}[t]
    \includegraphics[width=0.98\linewidth]{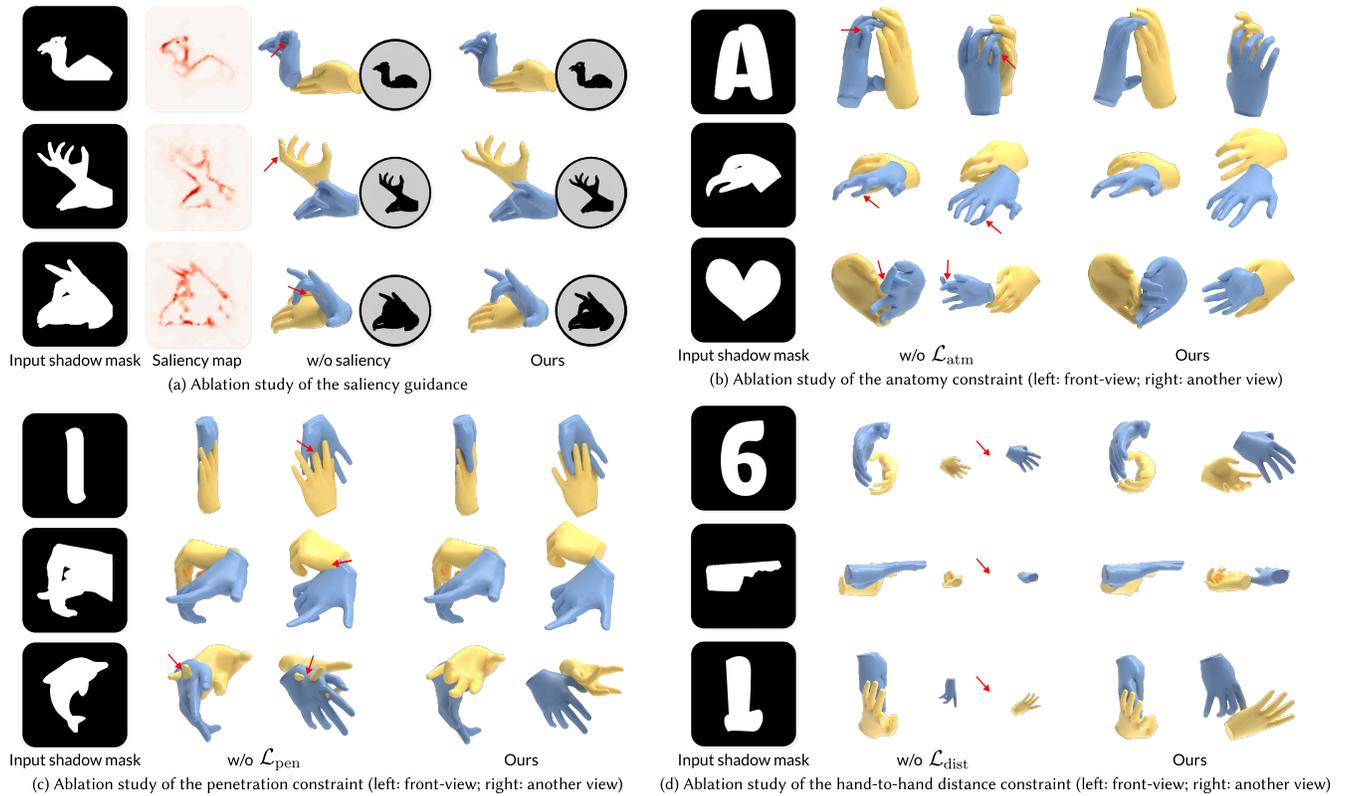}
     \caption{Ablation study of the key components in our shadow-feature-aware refinement module. }
    \Description[]{}
    \label{fig:ablation}
\end{figure*}

\paragraph{User study on quality comparison}
\label{para:user_study}
Procedure-wise, we showed each participant 60 sets (20 shapes, for each shape class) of results from the three baselines and our method.
%
The order of the results from different methods is randomly shuffled, with the associated target shape fixed on the left.
For each result, we asked the participant to rate it on \emph{Human-Global} and \emph{Human-Semantic} by comparing it with the target shape.
Table~\ref{tab:quantitative_results} (right) reports the average scores for each class.
For all three classes, our method consistently achieves better scores in \emph{Human-Global} and \emph{Human-Semantic} than the baselines, confirming the satisfying perceptual quality of our results.

\paragraph{User study on human demonstration}
In this study, the participants had to employ the physical setup described in Section~\ref{para:human_demo} to recreate six shadows using their hands: \textsc{Wolf}, \textsc{Sheep}, \textsc{Panther}, \textsc{Kangaroo}, \textsc{Dinosaur}, and \textsc{Stalin}.
In detail, we randomly and evenly split the ten participants in the first group into two sub-groups: 
the first sub-group aimed to reproduce the target shadows simply by taking the target shadow images as references, whereas the second sub-group performed the same task but additionally took our method's generated 3D hand models as references.
Then, the three judges in the second group scored the quality of the reproduced shadows, following the \emph{Human-Semantic} metric.
Besides, we recorded the time taken to reach the best hand shadow by watching the video recordings of the reproducing procedures.
Specifically, we recorded the timestamp of the best frame during the entire shadow-reproducing procedure for each shape within 3 minutes.  
Overall, our Hand-Shadow Poser helps reduce the average time taken by the participants, from 121.4 to 65.8 seconds, and improves the quality of the reproduced shadows, from 2.4 to 4.0 (on average).

\subsection{Model Analysis}
\paragraph{Ablation studies}
Beyond comparisons with baselines, we additionally conduct ablation on the shadow-feature-aware refinement module in our pipeline, including removing (i) the whole refinement module, (ii) the saliency guidance in similarity constraint, (iii) the anatomy constraint, (iv) the penetration constraint, and (v) the hand-to-hand distance constraint, from our full design.

For cases (i) and (ii), we first provide a quantitative analysis in Table~\ref{tab:ablation}. 
The~\emph{DINO-Semantic} score drops significantly after removing the refinement module, showing the importance of shadow-specific optimization in achieving fine-grained shadow alignment. 
Though the effect of the saliency guidance is not obvious in~\emph{LPIPS}, for which we speculate is insensitive to local characteristics, the~\emph{DINO-Global} and~\emph{CLIP-Semantic} both show a moderate performance degradation without the saliency map. 
Besides, given the same initial pose for refinement, the visual ablation in Figure~\ref{fig:ablation}~(a) clearly shows the impact of the saliency guidance in preserving prominent and intricate details, such as the eyes of \textsc{Camel} and \textsc{Donkey}.

As the remaining three constraints in cases (iii-v) are directly imposed on the 3D hand poses to aim for physical plausibility, we show their visual ablation results,~\ie, 3D hand poses in Figure~\ref{fig:ablation}~(b-d). 
Comparing the areas highlighted with the red arrows, we can observe severe physical artifacts of the hands, including poor anatomy (Figure~\ref{fig:ablation}~(b)), penetration (Figure~\ref{fig:ablation}~(c)), and excessive distance ((Figure~\ref{fig:ablation}~(d))), after removing each constraint, manifesting the effectiveness of each of the associated constraint.

\paragraph{Diversity analysis}
Further, we showcase multiple diverse results obtained by our method in Figure~\ref{fig:diversity}. 
For each case, we show four unique solutions, with the projected shadows closely resembling the input, while also remaining physically feasible.
This manifests the diversity with reasonable hand shapes introduced by our carefully-designed generative hand assignment module.

\paragraph{Robustness analysis}
Lastly, we study the robustness of our generalized hand-shadow alignment module. 
Figure~\ref{fig:hand_assignment_comparison} shows the coarse 3D hand poses of each hand without refinement (middle column in each case). 
For \emph{all assignment results}, the coarse hand poses exhibit contours, positions, and orientations that are approximately consistent with the input hand shapes, revealing our method's robustness to handle input masks of varying levels of uncertainty, particularly evident in the \textsc{Stanford bunny} case of \emph{Baseline 3}. 

\begin{figure}[t]
    \includegraphics[width=\linewidth]{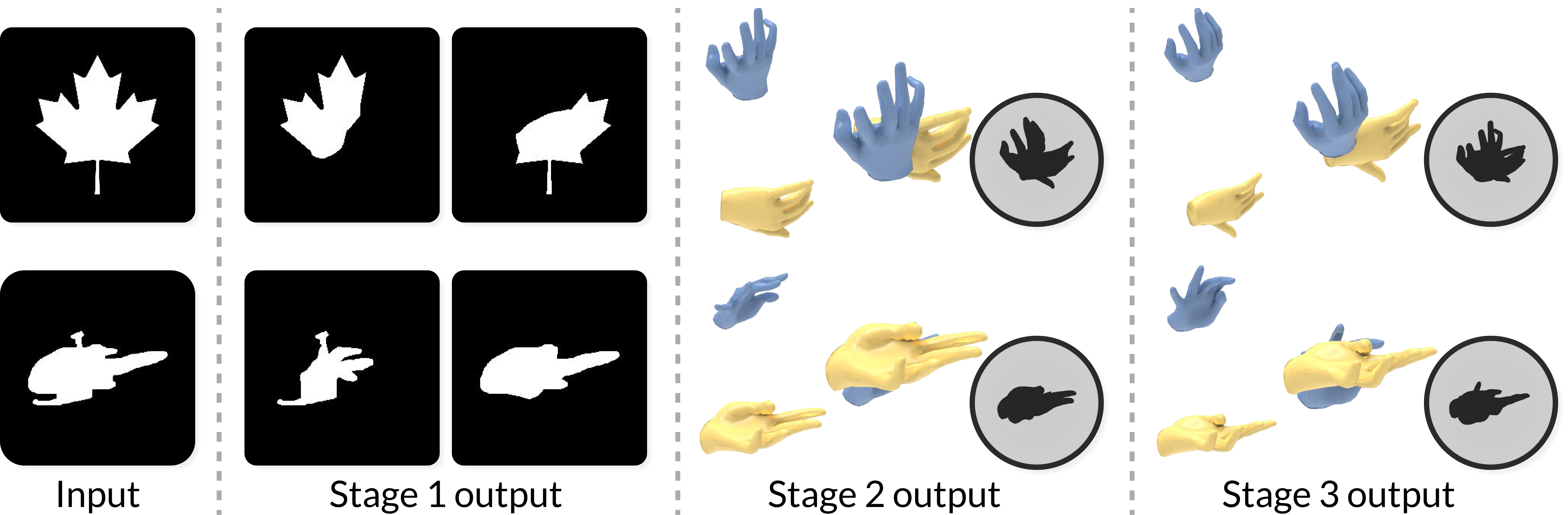}
    \vspace*{-4mm}
    \caption{
    Failure cases (\textsc{Maple} and \textsc{Chopper}). 
    Our method may not be able to work on arbitrary inputs with intricate details or thin structures.
    }
    \Description[]{}
    \label{fig:failure_cases}
\end{figure}

\paragraph{Failure case analysis}
\label{sec:failure_case}
Hand-Shadow Poser may not be able to produce reasonable results for arbitrary inputs, as not all shapes can be effectively reproduced by hands, particularly those with intricate details or thin structures; see Figure~\ref{fig:failure_cases}.
In such instances, our generative hand assignment may struggle to produce plausible hand shapes in Stage 1, thus leading to suboptimal hand reconstructions in Stage 2. 
Further, due to poor initialization, the inherent limitations of hand anatomy make it challenging for Stage 3 to refine poses to adequately fit the inputs.
\section{Conclusion, Limitations, and Future Works}
We presented the first comprehensive framework, namely Hand-Shadow Poser, to inversely create hand shadow arts from 2D shape inputs. 
We showcase the application of our approach to a wide variety of shapes, ranging from numbers and letters to classical hand shadows, and more challenging shapes of everyday objects.
We contribute three notable advances:
(i) first attempt to learn and reproduce hand shadow arts in a data-driven manner; 
(ii) a three-stage pipeline to decouple the anatomical constraints imposed by hand and semantic constraints imposed by shadow shape, with three novel components: generative hand assignment, generalized hand-shadow alignment, and shadow-feature-aware refinement. 
This decoupling design also frees us from building extensive domain-specific training data; and
(iii) an evaluation benchmark with a rich variety of shadow art samples of varying complexity, along with a family 
of metrics for quantitative assessment.
Also, we demonstrate the superior performance of our approach through extensive quantitative and qualitative comparisons with several alternative baselines and through user studies to evaluate human perception.
In the end, we performed a series of analyses, including ablation on key components, result diversity, and model robustness, to study the effectiveness of our proposed designs. 

Overall, the evaluation results highlight the generalizability and robustness of Hand-Shadow Poser in creating hand shadow arts with prominent features preserved for various types of input shapes, which can be further reproduced by human hands and 3D printing.

\paragraph{Limitations} 
The inverse hand-shadow-art problem is intriguing yet challenging.
Our work still has some limitations.
First, given an overly complicated shadow, such as shapes with small and thin structures, our approach may not be able to reproduce the shape due to the limited feasibility of human hands~(see Section~\ref{sec:failure_case}). 
Second, our approach cannot be directly applied to human hands of an arbitrary individual, due to variations in finger length and hand size, requiring customization by first specifying the hand-shape parameters of the individual.
Third, we assume a fixed light source, which might not be feasible for some target shapes that are formed by distorted shadows. 
Fourth, a critical challenge is to ensure the feasibility of humans, since not all two-hand poses are achievable due to the anatomical limits of the human body, which cannot be resolved merely through physical constraints on hands.
%
Last, our approach does not consider the forearm, which cannot be neglected in practice, as it may obscure the contour of the hand shadows.
To incorporate the forearm into our pipeline, a potential solution is to extend the MANO hand model with forearm parameters (\eg, via SMPL-X~\cite{smplx2019}) for pose optimization, with a penalty term in Stage 3 to deviate the forearm shadows from obstructing critical hand features. Additionally, the feasibility issue can be partially addressed by restricting the left/right hand swapping based on the arm position constraints.

\paragraph{Future Works}
Currently, our approach focuses on two hands to create hand shadow art from a single target shadow. 
First, we are interested in extending our approach to designing animated hand shadow arts for storytelling. 
Second, it would be interesting to incorporate hand-held object(s) in our approach, so that we may produce more intricate and visually appealing shadow results.
Lastly, given that some shadow plays involve more than two hands from multiple artists, it would be intriguing to adapt Hand-Shadow Poser to coordinate the hands of multiple humans, enabling more elaborate and captivating hand shadow art creation.
\begin{acks} 
This work was partially supported by the Research Grants Council of the Hong Kong Special Administrative Region, China, under Project T45-401/22-N and No. CUHK 14201321. 
Niloy J. Mitra was partially supported by gifts from Adobe and the UCL AI Centre.
Hao Xu thanks for the care and support from Yutong Zhang and his family.
\end{acks}

\bibliographystyle{ACM-Reference-Format}

\end{document}